\documentclass{eas}
\usepackage{graphicx,amsmath,amssymb,natbib,wasysym}
\bibpunct[]{(}{)}{;}{a}{}{,}
\newcommand{\D}[2]{\frac{{\rm d} #1}{{\rm d} #2}} 

\def\mnras{MNRAS}
\def\aap{A\&A}
\def\aaps{A\&AS}
\def\aj{AJ}
\def\apj{ApJ}
\def\apjs{ApJS}

\def\araa{ARA\&A}
\def\apjl{ApJ}
\def\pasj{PASJ}
\def\nat{Nature}
\def\icarus{Icarus}

\begin{document}

%%-----------------------------
%%      the top matter
%%-----------------------------
\title{Observation of rotation in star forming regions: clouds, cores, disks, 
and jets} 
\runningtitle{Observation of rotation in star forming regions}
\author{A. Belloche}\address{Max-Planck-Institut f\"ur Radioastronomie, Auf dem H\"ugel 69, D-53121 Bonn, Germany}
\begin{abstract}
Angular momentum plays a crucial role in the formation of stars and planets. 
It has long been 
noticed that parcels of gas in molecular clouds need to reduce their
specific angular momentum by 6 to 7 orders of magnitude to participate in the
building of a typical star like the Sun. Several physical processes on 
different scales and at different stages of evolution can contribute to this 
loss of angular momentum. In order to set constraints on these processes and
better understand this transfer of angular momentum, a detailed 
observational census and characterization of rotation at all stages of 
evolution and over all scales of star forming regions is necessary. This review 
presents the main results obtained in low-mass star forming 
regions over the past four decades in this field of research. It addresses the 
search and characterization of rotation in molecular clouds, prestellar and 
protostellar cores, circumstellar disks, and jets. Perspectives offered by
ALMA are briefly discussed.

\end{abstract}
\maketitle
%%-----------------------------
%%      your text
%%-----------------------------
%\pagebreak
\section{Introduction}
\label{s:intro}

Rotation is a ubiquitous phenomenon from the largest to the smallest scales in 
our Galaxy: the Galaxy itself rotates as a whole, its individual stars spin 
too -- like our Sun --, the majority of stars belong to binary systems, and the 
formation of planets around stars would not be 
possible if a certain amount of angular momentum was not present at the 
beginning of the star formation process. The angular momentum thus plays a 
significant role during the process of star and planet formation. The typical 
specific angular momenta measured on different scales and at different 
evolutionary stages, from dense cores in molecular clouds down to the Sun, are 
listed in Table~\ref{t:intro_joverm}. This table shows that a parcel of gas 
initially located in a dense core has to reduce its angular momentum by 6 to 7 
orders of magnitude in order to participate in the building of a typical star 
like our Sun. This puzzle has long been regarded as the 
``angular momentum problem'' in the field of star formation 
\citep[e.g.,][]{Spitzer78,Bodenheimer95,Mathieu04}.
Obviously, since angular momentum is a conserved quantity, this loss of 
angular momentum has to occur by transfer to other particles that will not be 
incorporated into the star. For instance, a fraction of the angular momentum 
initially present in a dense core can be stored by fragmentation into the 
orbital motion of a binary system. However, this fraction is typically on the 
order of a few percent only (see Table~\ref{t:intro_joverm}) and this process 
alone cannot solve the angular momentum problem. Magnetic braking may also 
play a significant role in carrying away angular momentum via Alfv{\'e}n waves 
during the early phases of star formation \citep[see, e.g.,][, and the review 
of P. Hennebelle in this volume]{Bodenheimer95}.

\begin{table}[t]
\begin{center}
\caption{Typical specific angular momenta from dense cores to the Sun}
\label{t:intro_joverm}
\vspace*{1ex}
\begin{tabular}{lcc}
    \hline
    \hline
    \noalign{\smallskip}
    \multicolumn{1}{c}{Object}        & $J/M$ & References \\
                                      & {\tiny (cm$^2$~s$^{-1}$)} & \\
    \hline
    \noalign{\smallskip}
    Dense cores in molecular clouds   & $10^{21-22}$ & 1 \\
    Protoplanetary disks              & $10^{19-21}$ & 2 \\
    Pre-main-sequence binaries        & $10^{19-20}$ & 3 \\
    Pre-main-sequence stars           & $10^{16-17}$ & 4 \\
    Extrasolar planetary systems (exoplanet(s) + star) & $10^{16-18}$ & 5, 6 \\
    Solar system (planets + Sun)      & $10^{17}$   & 7 \\
    Sun                               & $10^{15}$   & 8 \\
    \hline
\end{tabular}
\end{center}
References: 1: \citet{Goodman93}, 2: \citet{Williams11}, 3: \citet{Chen07},
4: \citet{Mathieu04}, 5: \citet{Armstrong07}, 6: \citet{Berget10}, 
7: \citet{Allen73}, 8: \citet{Pinto11}.
% Allen's astrophysical quantities 1973 (3rd version)
% M planetary system = 448 * 5.976e27 g
% J                  = 3.148e50 g cm-2 s-1
% J sun              = 1.63e48 g cm-2 s-1
% Pinto et al. 2011: J sun = 1.84e48 g cm-2 s-1
\end{table}

The specific angular momentum of a dense core has to be reduced by about two 
orders of magnitude during its gravitational collapse until the formation of a 
protoplanetary disk (see Table~\ref{t:intro_joverm}). During this protostellar 
phase, the ejection of matter in jets and outflows can partly contribute to 
this loss of angular momentum. Later on, the (proto)star-plus-disk system  has 
to reduce its specific angular momentum by still a few orders of magnitude to 
reach the level of our solar system, which appears to have a typical specific 
angular momentum compared to the currently known extrasolar planetary systems 
(see Table~\ref{t:intro_joverm}). Finally, the star itself has to
reduce its angular momentum by one to two orders of magnitude from the
pre main sequence to the main sequence at the age of the Sun. Star-disk 
interactions and magnetized stellar winds are thought to play a significant 
role in this respect (see the review of J. Bouvier in this volume). The 
solution of the angular momentum problem obviously involves different physical 
processes at various stages during the process of star formation. 

In order to better understand when and how these transfers of angular momentum 
occur during the process of star formation and set constraints on the physical
processes at work, it is essential to collect measurements of rotation on 
different scales and at different evolutionary stages of the star formation 
process. With this in mind, the aim of this review is to present an overview 
of the observational results that were obtained on rotation in star forming 
regions over the past four decades until February 2013. An earlier review, 
which also discussed the 
physical processes thought to be at work on a theoretical basis, was presented 
by \citet{Bodenheimer95} about two decades ago. Since then, many new 
observational results have been collected. These new results, as well as the 
earlier ones, are reviewed here critically. As a caveat, we stress that this 
review focuses on low-mass star forming regions only\footnote{Some aspects of 
rotation in high-mass star forming regions are discussed in, e.g., 
\citet{Pirogov03}, \citet{Cesaroni07}, \citet{Beltran11}, and \citet{Li12}.}.

The basic tools that can be used to probe rotation in star forming regions are 
presented in Sect.~\ref{s:tools}. Section~\ref{s:cores} addresses the search
for rotational signatures on large scales in molecular clouds and on smaller
scales in dense cores. Sections~\ref{s:envelopes} to \ref{s:jets} deal
with more evolved stages of star formation: rotation of protostellar 
envelopes is examined in Sect.~\ref{s:envelopes}, protoplanetary disks in 
Sect.~\ref{s:disks}, and jets and outflows in Sect.~\ref{s:jets}. The reader 
will find a short summary of the main results and ideas at the end of each 
section. Overall conclusions are presented in Sect.~\ref{s:conclusions} and 
some perspectives offered by the Atacama Large Millimeter/submillimeter Array 
(ALMA) are briefly described.

\section{How to probe rotation in star forming regions}
\label{s:tools}
 
This section describes basic tools that can be used to search for 
signatures of rotation in star forming regions.

\subsection{Centroid velocity}
\label{ss:vcent}

The centroid velocity of a molecular transition observed toward 
an astronomical object is the average velocity of the emitting particles
along the line of sight. It corresponds to the ``systemic'' velocity of the 
object for the line of sight toward its center. There are two main traditional 
ways to derive the centroid velocity. For a spectrum with a shape 
close to a Gaussian, a least-square fitting with the 3-parameter function
\begin{equation}
T_{\rm peak}\, {\rm e}^{-4 \ln(2) \frac{(v-v_{\rm g})^2}{FWHM^2}}
\end{equation}
yields the centroid velocity $v_g$, as well as the peak temperature 
$T_{\rm peak}$ and the linewidth $FWHM$. The second method consists in computing 
the first moment of the spectrum: 
\begin{equation}
v_{\rm f} = \frac{\sum\limits_{i=1}^N T_i \, v_i}{\sum\limits_{i=1}^N T_i}\,\,,
\end{equation}
with $N$ the number of channels in the selected velocity range over which
the emission is detected and $T_i$ the flux density in temperature scale in 
channel $i$. If the channels are independent, the uncertainty associated to 
the first moment is 
\begin{equation}
\sigma_f = \frac{\sigma_T}{\sum\limits_{i=1}^N T_i} \sqrt{\sum\limits_{i=1}^N (v_i-v_f)^2}\,\,,
\end{equation}
with $\sigma_T$ the rms noise level of the spectrum. The first method
can relatively easily handle multiple components along the line of sight but
convergence issues are encountered when the signal-to-noise ratio becomes too
low. The second method is mathematically simple and does not evolve any fitting 
process, but it is more difficult to deal with multiple components and
the equation may diverge when $\sum\limits_{i=1}^N T_i$ is close to zero,
for instance in case of a low signal-to-noise ratio.

\subsection{How does rotation manifest itself observationally?}
\label{ss:signatures}

\begin{figure}
\includegraphics[width=0.27\paperwidth,angle=0]{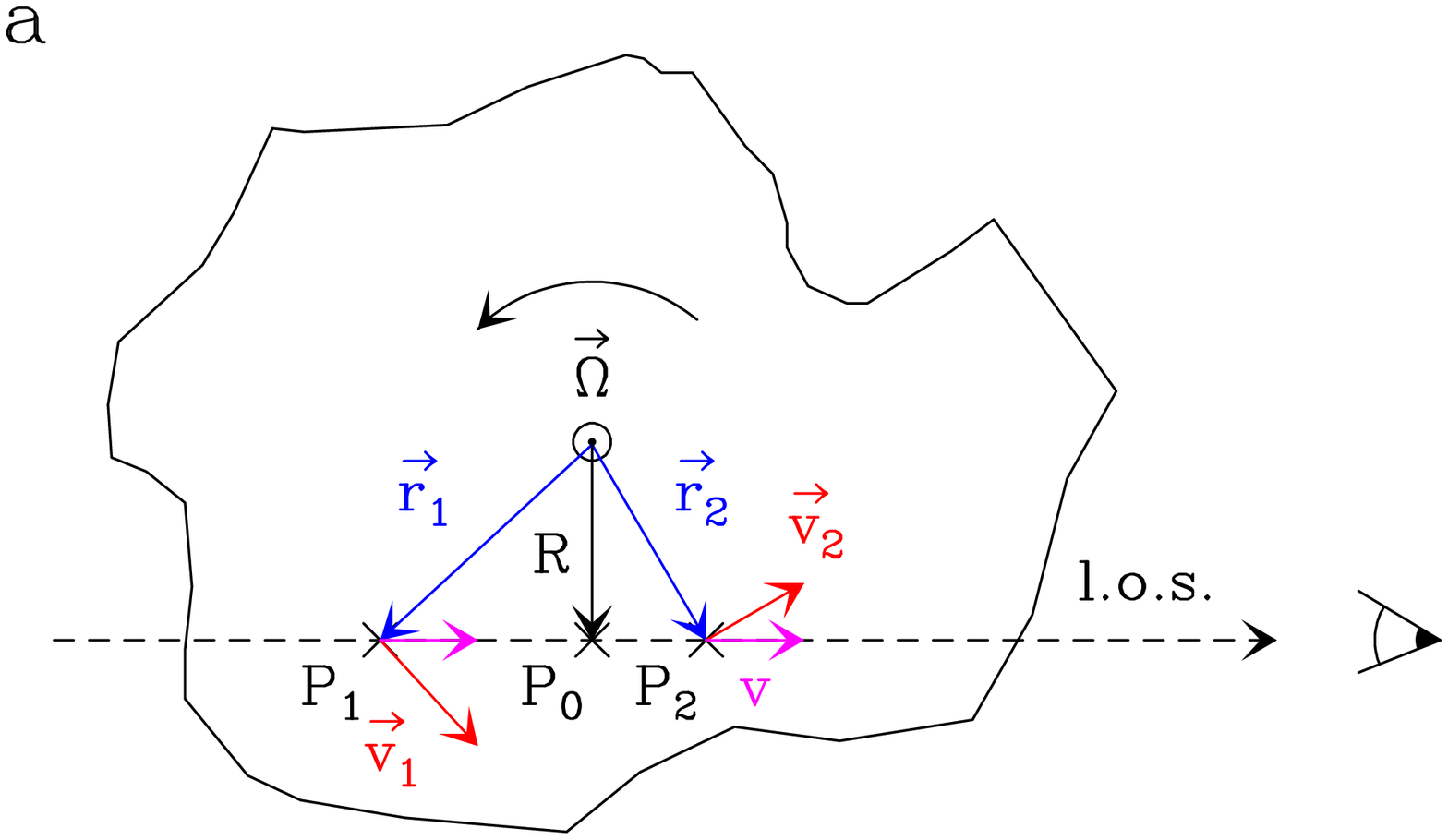}
\hspace*{2ex}
\includegraphics[width=0.27\paperwidth,angle=0]{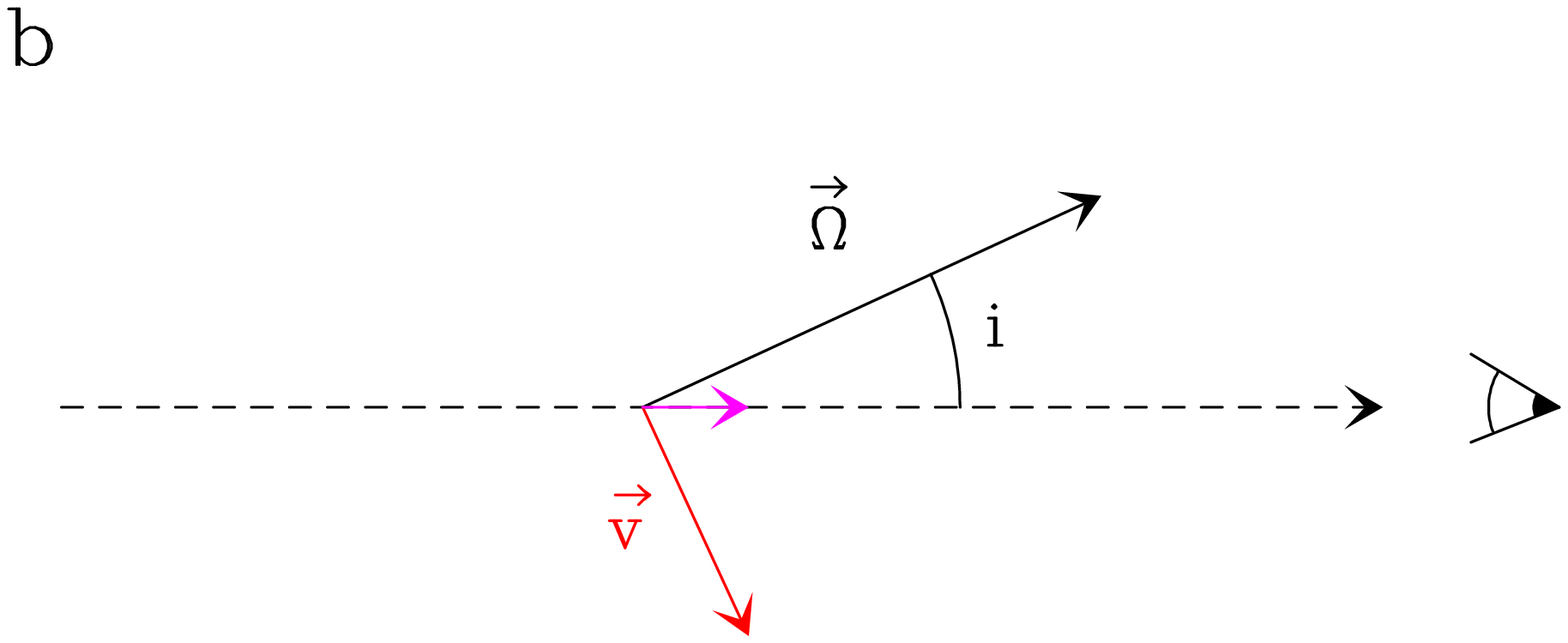}
\caption{\textbf{a} Projection of the rotation velocity of an object in 
solid-body rotation along the line of sight, for the particular case where the 
rotation axis is orthogonal to the line of sight, viewed from the top.
\textbf{b} Configuration when the rotation axis has an 
inclination $i$ to the line of sight, viewed from the side.}
\label{f:solidbody_drawing}
\end{figure}

Let's assume an object in solid-body rotation with an angular velocity 
$\vec{\Omega}$. Its rotation velocity field is defined by the equation 
$\vec{v} = \vec{\Omega} \times \vec{r}$. $\Omega$ being uniform for solid-body 
rotation, $v$ is proportional to $R$, the distance to the rotation axis. Let's 
take two points $P_1$ and $P_2$ along the line of sight, and let's call $P_0$ 
the point closest to the rotation axis along the line of sight and $R$ its
distance to the rotation axis (see 
Fig.~\ref{f:solidbody_drawing}a). The projection onto the line of sight of the 
velocity difference between $P_1$ and $P_2$ is
\begin{align}
v_{\rm diff}^{\rm los} & = \frac{1}{\mid\vec{r}_2-\vec{r}_1\mid} (\vec{r}_2-\vec{r}_1).(\vec{v}_2-\vec{v}_1)  \nonumber \\
                  & = \frac{1}{\mid\vec{r}_2-\vec{r}_1\mid} (\vec{r}_2-\vec{r}_1).(\vec{\Omega} \times \vec{r}_2 - \vec{\Omega} \times \vec{r}_1) \nonumber \\
                  & = -\frac{1}{\mid\vec{r}_2-\vec{r}_1\mid} \left(\vec{r}_1.(\vec{\Omega} \times \vec{r}_2) + \vec{r}_2.(\vec{\Omega} \times \vec{r}_1)\right) \nonumber \\
                  & = 0  \label{e:vdiff}
\end{align}
The projection onto the line of sight of the rotation velocity is thus 
independent of the position along the line of sight. This implies that the 
centroid velocity of the spectrum is equal to the rotation velocity of $P_0$
($v_{\rm cent} = \Omega R$) for a rotation axis 
orthogonal to the line of sight. The proof of Eq.~\ref{e:vdiff} is actually 
valid for any inclination $i$ of the rotation axis to the line of sight, and 
the centroid velocity becomes $v_{\rm cent}= \Omega R \sin i$ for the general 
case (see Fig.~\ref{f:solidbody_drawing}b). As a result, the velocity gradient
along the direction perpendicular to the projection onto the plane of the sky 
of the rotation axis is $\D{v_{\rm cent}}{R} = \Omega \sin i$ and is uniform
for solid-body rotation.

If the angular velocity depends on $R$ (but not on the azimuth), the object is
in differential rotation. For symmetry reasons, the centroid velocity for 
the line of sight with impact parameter $R$ still corresponds to the
rotation velocity at $P_0$ projected onto the line of sight.
Measuring the variations of centroid velocity as a function of $R$ thus
allows to derive $\Omega(R) \sin i$.

A search for rotation signature will start by investigating a map of centroid
velocity. A solid-body rotation will manifest itself as a uniform gradient of
centroid velocity, with a magnitude equal to $\Omega \sin i$. Estimating the
mean velocity gradient in such a map can thus be done by a simple planar
least-square fitting with the function 
$v_{\rm cent} = v_0 + a\Delta\alpha+b\Delta\beta$, with $\Delta\alpha$ and 
$\Delta\beta$ in radian \citep[][]{Goodman93}. The 
position angle of the direction of the mean velocity gradient is 
$PA = \tan^{-1} \left(\frac{a}{b}\right)$ and its magnitude is 
$\frac{\sqrt{a^2+b^2}}{d}$ ($= \Omega \sin i$), with $d$ the distance. The
magnitude is often given in km~s$^{-1}$~pc$^{-1}$ (1~km~s$^{-1}$~pc$^{-1}$ 
$= 3.2\,\,10^{-14}$~s$^{-1}$).

\subsection{Position-velocity diagrams}
\label{ss:pvdiag}

\begin{figure}
%\centerline{\includegraphics[width=0.28\paperwidth,angle=0]{Figs/posvit_mix_mod01_ees1209.eps}\hspace*{0.01\paperwidth}\includegraphics[width=0.225\paperwidth,angle=0]{Figs/posvit_mix_mod11_ees1209.eps}}
\centerline{\includegraphics[width=0.28\paperwidth,angle=0]{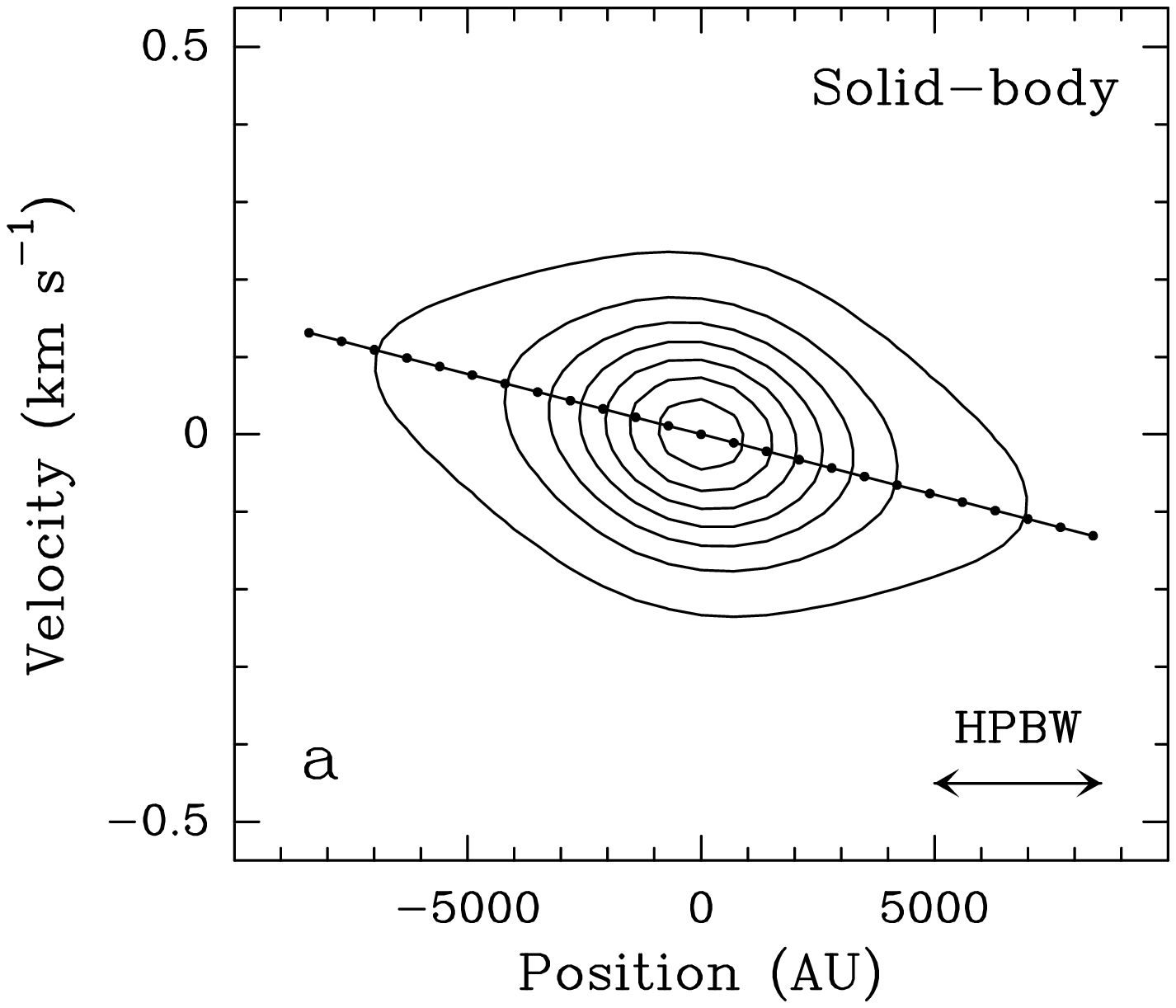}\hspace*{0.01\paperwidth}\includegraphics[width=0.225\paperwidth,angle=0]{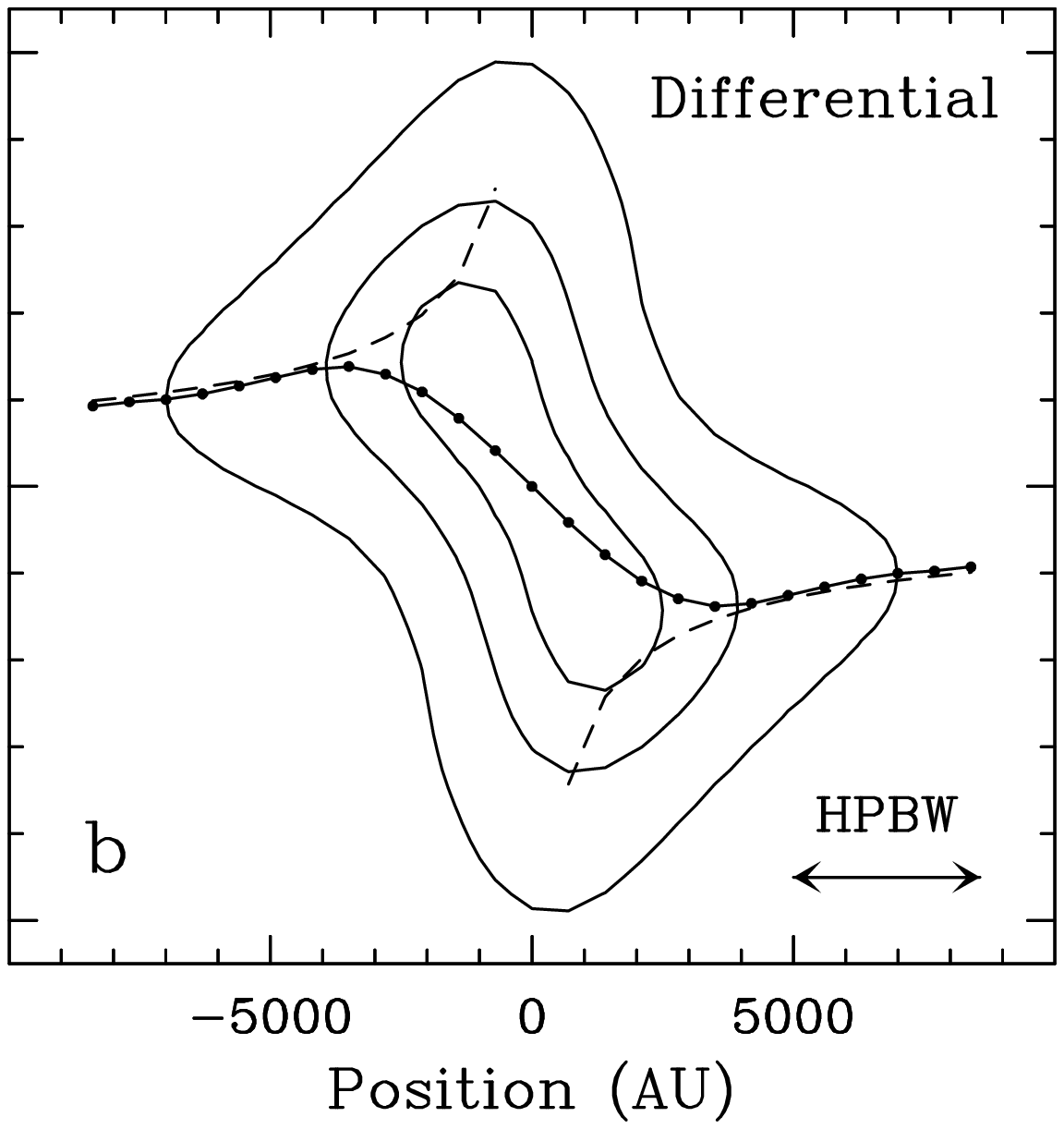}}
\caption{\textbf{a} Position-velocity diagram along the direction of the 
velocity gradient for a parametric model of a protostellar envelope in 
solid-body rotation with $\Omega = 4.2$~km~s$^{-1}$~pc$^{-1}$ and $i = 50^\circ$
\citep[adapted from][]{Belloche02b}. The emission (contours) was computed for 
C$^{34}$S 2--1 and a finite angular resolution ($HPBW = 3600$~AU). The curve 
marks the location of the centroid velocity at each position.
\textbf{b} Same as \textbf{a} but for an envelope in differential rotation 
with $\Omega = 12\,(R/3500~{\rm AU})^{-1.5}$~km~s$^{-1}$~pc$^{-1}$. The dashed 
line shows the expected variations of the centroid velocity for an infinite 
angular resolution.}
\label{f:pvdiag}
\end{figure}

Once the direction of the mean velocity gradient is known, it is instructive
to analyze the variations of the centroid velocity in a position-velocity (P-V)
diagram along this direction. Figure~\ref{f:pvdiag}a shows such a diagram
for a synthetic protostellar envelope in solid-body rotation. The centroid 
velocity curve is a straight line with a slope equal to the velocity gradient. 
The deviation from a straight line indicates differential rotation, as
illustrated in Fig.~\ref{f:pvdiag}b for an envelope with 
$\Omega \propto r^{-1.5}$. The shape of the centroid velocity curve
is affected by the finite angular resolution of the telescope in the
central parts, which prevents drawing any conclusion about the velocity field
within the inner region of diameter about twice the beam width
\hbox{($\sim 2\,HPBW$)}.
However, the emission contours in the P-V diagram provide additional 
information about the rotation velocity field even in the unresolved central 
parts. Fitting a synthetic contour map to the observed one allows to
derive constraints on this velocity field down to radii smaller than $HPBW$,
provided the contribution of other motions (infall, outflow, turbulence) is 
negligible or well known.

\subsection{Angular momentum and rotational energy}
\label{ss:janderot}

This section lists a number of useful equations for the case of an
object in solid-body rotation with a power-law density profile 
$\rho \propto r^{-\alpha}$ and a radius $R$ 
\citep[see, e.g.,][]{Goldsmith85}. The gradient of centroid velocity is 
$|\vec{\nabla}v_{\rm cent}| = \Omega \sin i$ (see Sect.~\ref{ss:signatures}). 
For a population of objects with random inclinations, the statistical 
correction to estimate the angular velocities consists in dividing the 
magnitudes of the velocity gradients by $\langle\sin i\rangle = \frac{\pi}{4}$.
The moment of inertia is 
$I = \frac{2}{3}MR^2\left(\frac{3-\alpha}{5-\alpha}\right)$, the angular
momentum $J = I\,\Omega$, and the specific angular momentum  
$j = \frac{J}{M} = \frac{I}{M}\,\Omega$. The rotational energy is 
$E_{\rm rot} = \frac{1}{2}I\,\Omega^2$ and the gravitational energy is:
\begin{equation}
E_{\rm grav} = -\frac{GM^2}{R}\left(\frac{3-\alpha}{5-2\alpha}\right).
\end{equation}
The ratio of rotational to gravitational energies is:
\begin{equation}
\beta_{\rm rot} = \frac{E_{\rm rot}}{|E_{\rm grav}|} = 
    \frac{1}{3}\frac{R^3\Omega^2}{GM}\left(\frac{5-2\alpha}{5-\alpha}\right).
\end{equation}
For a uniform density ($\alpha = 0$), we obtain  $I = \frac{2}{5} MR^2$, 
$j = \frac{2}{5}R^2\Omega$, and 
$\beta_{\rm rot} = \frac{1}{3}\frac{R^3\Omega^2}{GM}$. For 
a centrally-peaked profile with $\alpha = 2$, $I = \frac{2}{9} MR^2$, 
$j = \frac{2}{9}R^2\Omega$, and 
$\beta_{\rm rot} = \frac{1}{9}\frac{R^3\Omega^2}{GM}$.
Note that $\beta_{\rm rot}$ is proportional
to $\frac{\Omega^2}{\rho_{\rm mean}}$, with $\rho_{\rm mean}$ the mean density.

\subsection{Basic procedure to reveal the presence of rotation}
\label{ss:tools_summary}

The basic procedure to reveal the presence of rotation requires a tracer being
optically thin in order to trace all material along each line of sight. 
Of course, only the material at densities (roughly) higher than the critical 
density of the molecular transition is probed. One should in addition 
keep in mind that the molecule abundance may not be uniform due to gas-phase 
chemical reactions and/or gas depletion/release onto/from the dust grain 
surface. This can 
affect the derivation of the centroid velocity. The advantage of an optically 
thin tracer is also that its lineshape will not be distorted by optical depth 
effects.

The first step of the procedure is to search for a one-dimensional, systematic
pattern in a map of centroid velocity. The second step consists in deriving
the direction of the mean velocity gradient (via, e.g., planar least-square 
fitting). The final step is to analyze the P-V diagram along the direction
of the mean velocity gradient. The curve of centroid velocity will indicate
if the object is in solid-body or differential rotation depending if it
is a straight line or not. In addition, comparing the emission contours 
to synthetic P-V diagrams can yield constraints on the rotation velocity
field at scales even smaller than the beam.

This procedure is relatively simple but one should keep in mind that,
depending on the geometry of the source, other types of motion, such as 
infall, outflow, or shear motions can also produce velocity gradients that
could be mistaken as due to rotation.

\section{Rotation of molecular clouds and prestellar cores}
\label{s:cores}

This section addresses the search for rotation in the prestellar phase,
from the scales of molecular clouds down to prestellar cores.

\subsection{Rotation of molecular clouds}
\label{ss:clouds}

Velocity gradients in tracers such as CO and $^{13}$CO in emission or H$_2$CO 
in absorption were found in large-scale maps of molecular clouds in the 70's. 
They were often interpreted as indications of rotation. \citet{Fleck81} 
compiled such measurements of 13 ``rotating'' clouds. Their angular 
velocities $\Omega$ normalized to the galactic angular velocity 
$\Omega_{\rm G}$ ranged from 7 to 300. The early interpretation of these high
values was that the angular momentum of these clouds was acquired from 
galactic rotation, starting with corotation and then spin-up during cloud 
contraction. However, this interpretation was questioned by the random 
orientation of rotational axes of early-type stars, binaries, and molecular 
clouds with respect to $\vec{\Omega}_{\rm G}$. For the cloud sample mentioned
above, $\Omega(R) \propto R^{p}$, with $R$ the cloud radius at which the 
velocity gradient was measured and $p \sim -\frac{2}{3}$ 
\citep[see Fig.~1 of][]{Fleck81}. \citet{Fleck81} interpreted this
relation as well as the cloud size distribution as resulting from 
the turbulent properties of the interstellar medium, a Kolmogorov cascade with 
turbulence being sustained on large scales by the shearing action of 
differential galactic rotation. Cloud ``rotation'' would then originate in 
turbulence vorticity.

\citet{Goldsmith85} analyzed a sample of 16 clouds without evident signs of
high-mass star formation. The sample was biased to include only ``rotating''
clouds, i.e. clouds with a clear velocity gradient. The cloud sizes range
from 0.1 to 17~pc with a median of 0.6~pc, and the typical densities are
$n_{{\rm H}_2} \sim 10^3$~cm$^{-2}$. Under the assumption of solid-body
rotation, they found specific angular momenta scaling as $R^{1.4}$, i.e.
$\Omega \propto R^{-0.6}$, consistent with the findings of \citet{Fleck81}.
They interpreted this relation as evidence for loss of angular momentum in
the process of cloud contraction and fragmentation. They suggested that this 
loss could be due to the redistribution of angular momentum in the orbital
motions of fragments or to magnetic braking that transfers angular momentum to 
larger scales. Under the assumption of uniform density, they derived 
$\beta_{\rm rot}$ ranging from 0.04 to 2.5, with a median of 0.25, suggesting 
that rotation is significant but does not dominate the cloud energetics. 
All the conclusions are based on the assumption that the velocity gradients do 
trace rotation. Examining in more details the $^{13}$CO~1--0 P-V diagrams of 
these clouds, the interpretation appears in some cases ambiguous: B361 has
a P-V diagram suggestive of differential rotation along the axis of its 
mean velocity gradient \citep[see Fig.~3 of][]{Arquilla85}, but the velocity 
dispersion along the orthogonal axis is as large as the velocity variations 
along the axis of the velocity gradient \citep[see Fig.~3b of][]{Arquilla86}.

Finally, under the assumption that linear velocity gradients observed in their
sample of 5 giant molecular clouds are due to rotation, \citet{Imara11} find 
that the specific angular momentum of these clouds is smaller than that in the
surrounding atomic gas out of which these clouds formed. Even more 
importantly, they find that the velocity gradient position angles in the 
molecular and atomic gas are largely divergent, which leads them to suggest 
that rotation may not be the best explanation of the velocity fields observed 
in giant molecular clouds.

Overall, all these results suggest that the interpretation of velocity 
gradients on molecular cloud scales as due to rotation is by no means 
straightforward and robust 
\citep[but see][ for an alternative view]{Phillips99}.

\subsection{Rotation of dense cores}
\label{ss:cores}

The presence of rotation on smaller scales corresponding to dense cores
traced with ammonia ($n_{{\rm H}_2} \sim 10^4$~cm$^{-2}$) was investigated by 
\citet{Goodman93}. Their sample contains 43 cores with radii ranging from 0.1 
to 0.5~pc. Significant velocity gradients were found in a large fraction of the
sample. Some neighboring cores have non-parallel velocity gradients (see, 
e.g., their Fig.~4a). The gradient directions do not correlate with the core 
elongations. The orientations of the gradients are random within a complex,
for instance for the subsample of 16 cores in Taurus. At least two cores,
L1251A and L1251E, have a rotation axis not parallel to the magnetic field.
Finally, assuming uniform density and not correcting for inclination, 
$\beta_{\rm rot}$ is less than 10\% for most cores, with a median of 0.03, 
pointing to a minor role of rotation in supporting the dense cores, in 
agreement with the absence of correlation between rotation axis and direction 
of elongation.

The angular momenta derived by \citet{Goodman93} correlate with the radii of 
the NH$_3$ dense cores as $R^{1.6}$, in a very similar way as for the
molecular clouds (Sect.~\ref{ss:clouds}). However, the authors pointed out 
that, given $j \propto \Omega R^2$, this correlation is dominated by $R^2$. 
The correlation of the angular velocities with radius, 
$\Omega \propto R^{-0.4}$, is actually very weak for their sample. In addition, 
they did not find any correlation between $\beta_{\rm rot}$ and $R$, which could 
result from $\beta_{\rm rot}$ being proportional to $\Omega^2/\rho_{\rm mean}$, 
$\Omega$ not correlating (or only weakly) with $R$, and ammonia probing a 
small dynamic range of densities.

\citet{Caselli02} probed rotation of dense cores at even higher densities with 
N$_2$H$^+$ 1--0 ($n_{{\rm H}_2} \gtrsim 10^5$~cm$^{-3}$). Their sample includes
57 cores, all previously mapped in ammonia. On average, the angular velocity
derived with N$_2$H$^+$ is 1.6 times larger than with NH$_3$, but with a large
dispersion (1.0). The directions of the N$_2$H$^+$ and NH$_3$ gradients are
well correlated, and the distributions of $\beta_{\rm rot}$ are very similar for 
both tracers ($\langle\beta_{\rm rot}\rangle = 0.02$ and 0.03, respectively). 
The planar least-square fitting method was also applied on subregions of each 
core. This revealed that many cores have internal variations of magnitude and
direction of velocity gradient. Simple solid-body rotation thus appears to be 
rare at the scale of dense cores.

\subsection{Probing higher angular resolution}
\label{ss:cores_highres}

L183 is a centrally condensed, chemically evolved, high-column density 
prestellar core. Recent interferometric observations in N$_2$H$^+$~1--0 
revealed the presence of a velocity gradient on small scales (3300~AU),
about 5 times larger than and with the same direction as the gradient measured
on larger scales (10300~AU) with single-dish telescopes \citep[][]{Kirk09}. The 
authors interpreted the gradients as due to rotation and concluded that the 
core inner parts span up with rough conservation of specific angular momentum.
The direction of the velocity gradient was claimed to be roughly
perpendicular to the projection onto the plane of the sky of the magnetic
field traced by dust polarization, possibly indicating an important role of 
the magnetic field in the core evolution. However, there are two caveats to this
result: the velocity gradient is seen only in the northern part of the core, not
toward the main peak traced with N$_2$H$^+$, and the positions of the 
continuum and N$_2$H$^+$ peaks do not match, suggesting that the N$_2$H$^+$ 
abundance is not uniform, which may affect the measurements of centroid 
velocities.

\subsection{Complex motions in protoclusters}
\label{ss:cores_protoclusters}

The previous sections addressed the search for rotation in relatively isolated 
dense cores. Since many stars form in clusters, it is relevant to investigate
the rotational properties of prestellar cores located in protoclusters.
The protocluster L1688 in Ophiuchus consists of 6 dense clumps, Oph~A to F,
within which 60 prestellar cores are embedded \citep[][]{Motte98}. Forty-eight 
of them were surveyed in N$_2$H$^+$ 1--0 \citep[][]{Andre07}. The centroid
velocities of the cores trace their motions within the protocluster. They
reveal a large-scale ($\sim 1.1$~pc) velocity gradient of magnitude 
$\sim 1.1$~km~s$^{-1}$~pc$^{-1}$ across the cloud  
\citep[see Fig.~7 of][]{Andre07}. The position angle (east from north) of the 
gradient changes from 117$^\circ$ to 186$^{\circ}$ when the cores in
Oph~B are excluded from the fit. It is thus unclear whether this large-scale
velocity gradient really traces rotation of the protocluster. The N$_2$H$^+$
centroid-velocity map of each dense clump reveals its own kinematics 
\citep[see Fig.~6 of][]{Andre07}. On the one hand, the velocity 
structure within Oph~A, B1, and B2 is highly complex, with no clear 
one-dimensional velocity pattern. It is thus difficult to conclude anything 
about rotation in these clumps. On the other hand, simple velocity gradients
in Oph~C and E are consistent with the presence of rotation. In Oph~F,
two unrelated components seem to overlap along the line of sight. Each 
component has a simple velocity gradient, possibly due to rotation. 

\begin{figure}
%\centerline{\includegraphics[width=0.22\paperwidth,angle=270]{Figs/oph_j2m.eps}}
\centerline{\includegraphics[width=0.22\paperwidth,angle=270]{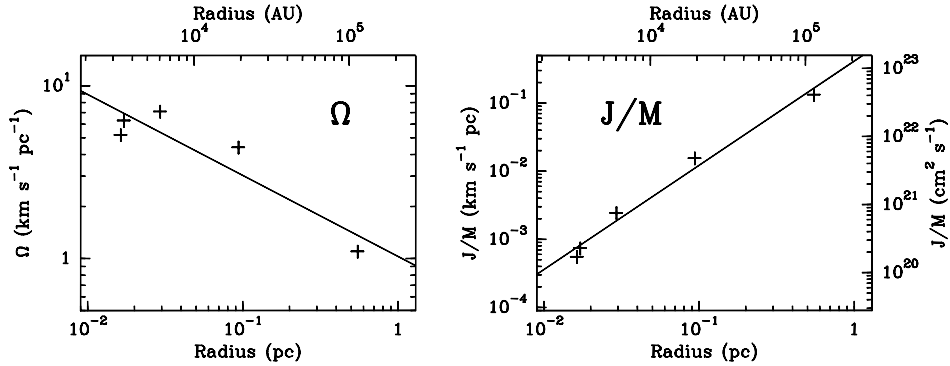}}
\caption{\textbf{a} Angular velocity as a function of radius for the 5 
structures in L1688 with clear velocity gradient: L1688 as a whole, Oph~C-S, 
E-MM2d, E-MM4, and F-MM1, based on \citet{Belloche02b}. \textbf{b} Specific 
angular momentum as a function of radius for the same structures as in 
\textbf{a}. No correction for inclination was applied and solid-body rotation 
and uniform density were assumed. In each panel, the solid line is the result 
of a least-square linear fitting in logarithmic scales. The slopes are $-0.5$ 
and 1.5, respectively.}
\label{f:l1688}
\end{figure}

The prestellar cores embedded in the clumps with possible signs of rotation 
represent only $\sim 25\%$ of the full N$_2$H$^+$ sample of L1688. It is thus 
difficult to draw general conclusions about the rotational properties of
prestellar cores in protoclusters. However, a closer look at the kinematics in 
Oph~E, which was also mapped in H$^{13}$CO$^+$~1--0, DCO$^+$~2--1, and 
DCO$^+$~3--2 by \citet{Andre07}, is instructive. The mean velocity gradient 
over the clump has a magnitude of $5.2 \pm 0.3$~km~s$^{-1}$~pc$^{-1}$ at 
$PA = -108^\circ \pm 4^\circ$. Fitting the two embedded prestellar cores 
Oph~E-MM2d and E-MM4 separately yields velocity gradients of magnitude 
$5.2 \pm 0.3$~km~s$^{-1}$~pc$^{-1}$ at $PA = -126^\circ \pm 15^\circ$ and 
$7.1 \pm 0.4$~km~s$^{-1}$~pc$^{-1}$ at $PA = -48^\circ \pm 8^\circ$, respectively 
\citep[see Fig.~5.15 of][]{Belloche02b}. The velocity gradients around 
Oph~E-MM2d and E-MM4 are significantly not parallel. If they are due to 
rotation, then their misalignment likely results from the turbulent nature of
rotation in protoclusters. This extends to much smaller scales the 
conclusion on the significant role of turbulence drawn on cloud scales by 
\citet{Fleck81} (see Sect.~\ref{ss:clouds}). This conclusion is further 
strengthened by the scaling of the angular velocity as a function of radius. 
We find $\Omega \propto R^{-0.5}$ and $j \propto R^{1.5}$ for the five 
structures with clear velocity gradient (see Fig.~\ref{f:l1688}). Even if the 
statistics is poor, this tentative correlation of $\Omega$ with $R$ in L1688 is 
close to the correlation derived on larger scales for a sample of clouds by 
\citet{Fleck81} which was interpreted as a signature of interstellar 
turbulence.

Assuming solid-body rotation and no correction for inclination, the velocity 
gradients measured for Oph~C-S, E-MM2D, E-MM4, and F-MM1 imply 
$\beta_{\rm rot}$ ranging from 2 to 22\% if the density is uniform, or from 0.7 
to 7\% if $\rho \propto r^{-2}$ \citep[see Table 5.7 of][]{Belloche02b}. Like 
for the isolated cores mentioned in Sect.~\ref{ss:cores}, rotation is 
energetically not dominant in the protocluster L1688, down to scales of a few 
1000~AU. Finally, we note that the rotation period associated with the derived 
angular velocities of the prestellar cores in L1688 is on the order of 
$10^6$~yr without correction for inclination. This is a factor 2 to 10 longer 
than the lifetime of the L1688 prestellar cores estimated by \citet{Andre07}. 
This confirms that rotation is dynamically not dominant at the scale of 
prestellar cores in this protocluster.

\subsection{Peculiar motions in starless cores}
\label{ss:cores_peculiar}

This section addresses two cases further illustrating the difficulties in
probing rotation in starless cores. L1506C is a large, low-density, starless 
core embedded in a filament in Taurus. Its inner parts ($r < 0.15$~pc) are 
characterized by a very low level of turbulence 
($\sigma_{\rm turb} < 47$~m~s$^{-1}$) and a high level of C$^{18}$O depletion, 
unexpected given the low density of the core \citep[][]{Pagani10}. The authors 
proposed that this high level of depletion is related to the low level of 
turbulence promoting dust coagulation, which in turn may decrease the 
desorption efficiency. An even more puzzling property of this core is its 
kinematical structure: \citet{Pagani10} report a velocity gradient in 
$^{13}$CO with a direction \textit{opposite} to the velocity gradients traced by 
C$^{18}$O and N$_2$H$^+$. Their detailed radiative transfer analysis shows that 
the inner parts are contracting ($v_{\rm inf} = 0.11$~km~s$^{-1}$) and rotating 
with $\Omega \propto r^{-1.5}$, and the outer parts are expanding 
($v_{\rm exp} = 0.09$~km~s$^{-1}$) and rotating in opposite direction. 
\citet{Pagani10} interprete these properties as evidence for decoupling of the 
inner core from the external parts, and suggest that L1506C is a prestellar 
core in the making. The peculiar velocity fields (infall/expansion, 
counter-rotation) remain however to be explained physically: the authors 
suggest that they may result from oscillations produced by magnetic torques. 
Alternatively, the velocity gradients may not trace rotation.

The prototypical, isolated, starless core B68 is supported by thermal pressure
\citep[][]{Alves01,Lada03}. \citet{Lada03} reported an approximately east-west 
velocity gradient, which they interpreted as rotation with 
$\beta_{\rm rot} \sim 4\%$. However, the presence of a ``bullet'' to the 
south-east of B68 led \citet{Burkert09} to propose that B68 is undergoing a 
collision with this small core, which they qualitatively reproduced with 
hydrodynamic simulations. This scenario was recently followed-up by 
\citet{Nielbock12} who analyze larger maps of centroid velocity. They 
establish a connection between the systemic velocity of the putative colliding 
small core and the velocity gradient seen across B68 (see their Fig.~17). On 
even larger scales, their $^{13}$CO 2--1 centroid-velocity map shows that the 
velocity structure of B68 is related to a larger-scale gradient along a 
large-scale underlying filamentary structure (see their Fig.~18). This 
velocity gradient could be due to streaming motions of cores along the 
filamentary structure rather than rotation. This would support the collision 
interpretation for B68.

\subsection{Reliability of angular momenta derived observationally}
\label{ss:j2d_reliability}

The rotation measurements reported in the previous sections are based on 
two-dimensional centroid-velocity maps. The loss of the third 
spatial dimension inherent in astronomical observations 
may bias the interpretation of velocity gradients, and in particular the
calculation of angular momenta. \citet{Dib10} investigate this issue by
means of three-dimensional magnetohydrodynamic simulations of isothermal,
self-gravitating clouds with decaying turbulence. The rms Mach number of
turbulence in their simulations is similar to those of the Ophiuchus (L1688) 
and Perseus molecular clouds. The dense cores generated in these simulations
show a variety of morphologies, from roundish to filamentary (see their 
Fig.~3). The authors compute projected maps of centroid velocity to compare
the outcome of the simulations to observations. As their Fig.~9 shows,
turbulence generates velocity gradients qualitatively similar to those 
measured in molecular cloud cores. \citet{Dib10} compute the distribution of 
3D specific angular momenta ($j_{\rm 3D}$) of their synthetic magnetized cores.
They find a median about 5 to 10 times lower than the median of angular 
momenta ($j_{\rm 2D, obs}$) derived observationally from two-dimensional 
centroid-velocity maps of NH$_3$ and N$_2$H$^+$ dense cores (see 
Sect.~\ref{ss:cores}). 

To understand this discrepancy, they also compute the 
specific angular momenta of the synthetic cores by following the observational
procedure ($j_{2\rm D}$), i.e. measuring the mean velocity gradient in projected
maps of centroid velocity and assuming solid-body rotation and uniform 
density to compute the angular momentum. They find $j_{\rm 2D}$ one order of 
magnitude larger than $j_{\rm 3D}$ for their synthetic cores, i.e. $j_{\rm 2D}$
values consistent with the observed ones ($j_{\rm 2D,obs}$). The ``rotational'' 
properties of the synthetic cores are thus very similar to the observed ones, 
but this analysis suggests that the assumption of solid-body rotation and 
uniform density to compute the specific angular momentum of a prestellar core 
based on a two-dimensional map overestimates the true value of the angular 
momentum. Assuming centrally peaked density profiles is not sufficient to 
remove the discrepancy. Interestingly, \citet{Dib10} find a correlation between 
$j_{\rm 3D}$ and $R_{\rm 3D}$ somewhat steeper than between $j_{\rm 2D}$ and 
$R_{\rm 2D}$ (exponent 1.8--2 vs. 1.1--1.3).

A puzzling result of this work, however, is that the synthetic and observed
cores have similar distributions of $\beta_{\rm rot}$ ($\beta_{\rm rot, 3D}$ for 
the former and $\beta_{\rm rot, 2D, obs}$ for the latter) while 
$\beta_{\rm rot, 3D} < \beta_{\rm rot, 2D, obs}$ could have been expected 
\textit{a priori} given that $j_{\rm rot, 3D} < j_{\rm rot, 2D, obs}$. A key 
result to solve this puzzle is provided by \citet{Offner08} who did a similar 
analysis for a population of non-magnetized synthetic cores. They obtain 
$\frac{j_{\rm 3D}}{j_{\rm 2D}} \sim 0.01$--0.1 for the synthetic cores (see 
their Fig.~5), in rough agreement with the findings of \citet{Dib10}. In 
addition, they compute $\frac{\beta_{\rm rot, 3D}}{\beta_{\rm rot,2D}}$ for the 
synthetic cores and obtain a median value close to 1, which explains why 
\citet{Dib10} find a good agreement between $\beta_{\rm rot, 3D}$ and 
$\beta_{\rm rot, 2D, obs}$. The reason why 
$\frac{\beta_{\rm rot, 3D}}{\beta_{\rm rot,2D}} \sim 1$ while 
$\frac{j_{\rm 3D}}{j_{\rm 2D}} \sim 0.01$--0.1 is unclear. We speculate that it 
is due
to the scalar nature of $\beta_{\rm rot, 3D}$ as opposed to the vectorial nature 
of the angular momentum. Two parcels of gas with opposite $\vec{J}$ will not 
contribute to the total angular momentum while they will both contribute to 
the ``rotational'' energy. This illustrates that $\beta_{\rm rot}$ certainly 
includes more than pure rotational energy.

\subsection{Summary}
\label{ss:cores_summary}

Given their scaling properties, the velocity gradients measured in molecular 
clouds and interpreted as rotation likely originate in turbulence vorticity.
At the scale of dense cores, the rotational energy is only a few percent of the
gravitational energy. Rotation is thus energetically not dominant. The direction
of rotation axis is not correlated with the core elongation or the
magnetic-field direction. The correlation between the specific angular 
momentum and the radius ($j \propto R^{1.4}$) is similar to the one found on 
larger scales for molecular clouds. Interpreted in terms of rotation, it
suggests a loss of angular momentum during the contraction process of dense
cores. This loss could be due to magnetic braking, gravitational torques, or
result from the transfer of angular momentum into the orbital motion of 
fragments. However, as for clouds, this behaviour may simply result from
the properties of interstellar turbulence. The interpretation of
velocity gradients in clouds and dense cores as rotation is therefore not 
unique. In any case, although often used as a zeroth-order approximation, 
solid-body rotation appears to be rare, which, in combination with the 
two-dimensional nature of astronomical data, likely results in overestimating 
the specific angular momenta and the ratios of rotational to gravitational 
energies. Finally, the velocity structure in protoclusters or ``non-isolated'' 
cores is often complex, resulting from, e.g., confusion along the line of 
sight, shear motions, or streaming motions along filamentary structures. 
In protoclusters, turbulence seems again to play a major role in the 
production of velocity gradients. Probing rotation at the prestellar stage
is thus often more challenging than one could naively expect.

\section{Rotation of protostellar envelopes}
\label{s:envelopes}

\subsection{General properties of protostars}
\label{ss:protostars}

Protostars are systems in the main accretion phase, with a stellar embryo -- 
i.e. an hydrostatic object -- at their center, accumulating mass from a 
collapsing envelope and/or accreting circumstellar disk. This accretion 
process is systematically accompanied by ejection of matter in the form of 
bipolar jets and outflows. Two types of protostars have been defined 
observationally: Class~0 protostars have most of their mass still in the 
envelope ($M_{\rm env} > M_\star$) while Class~I protostars have accreted most of 
their mass ($M_{\rm env} < M_\star$) \citep[][]{Andre93,Andre00}. In the Class~0 
phase, the envelope is still prominent and should retain memory of its initial 
conditions, and in particular have similar properties as in the prestellar
phase. In the Class I phase, the system is dominated by a star-disk system,
with a residual envelope.

Studying protostars is highly relevant for understanding star formation, and 
in particular the evolution of angular momentum, because a protostar 
\textit{will} form a star. This is not necessarily the case for starless 
cores, some of them being maybe only transient structures 
\citep[see, e.g.,][ for a discussion of the fate of starless cores]{Belloche11}.
The advantage of protostars over starless cores for searching for rotation is
that the position of the system ``center'' is known. In addition, jets and 
outflows, which are thought to be driven by magnetocentrifugal 
acceleration \citep[e.g.,][]{Konigl00,Shu00}, define a ``natural'' axis for 
rotation \textit{a priori}. In addition, they can give clues about the 
inclination of the system \citep{Cabrit90a}, which is needed to derive the
angular velocity and angular momentum.

\subsection{Rotation and outflow axis}
\label{ss:axes}

As stated in Sect.~\ref{ss:protostars}, we naively expect to find velocity 
gradients tracing rotation in protostellar envelopes in the direction 
perpendicular to the outflow axis. In practice, this is not often the case. 
\citet{Curtis11} surveyed starless and protostellar cores in Perseus in 
C$^{18}$O 3--2. Seven Class~0 cores have a simple bipolar CO outflow and a 
significant C$^{18}$O velocity gradient. For two of them, the velocity gradient 
is roughly orthogonal to the outflow axis, as we expect for rotation, but for 
5 of them, it is roughly parallel. This may indicate that the rotation axis 
changes direction from large scales ($n_{{\rm H}_2} \sim 10^4$--$10^5$~cm$^{-3}$) 
to small scales where the jet/outflow is launched 
\citep[at most a few AU, see Sect.~\ref{s:jets} and, e.g.,][]{Ferreira06}. 
Alternatively, C$^{18}$O 3--2 observed with single-dish telescopes may 
be a poor probe of envelope rotation. C$^{18}$O is indeed known to suffer from 
depletion at high density. Entrainment by the outflow may in addition 
contaminate the P-V diagrams. In high-density protoclusters, C$^{18}$O 3--2 may 
also be too sensitive to the ambient cloud.

\citet{Tobin11} surveyed 17 isolated, mostly Class~0 but also a few Class~I, 
protostellar envelopes in N$_2$H$^+$~1--0 and NH$_3$~(1,1) with 
single-dish telescopes and interferometers. The single-dish maps reveal 11 
sources (out of 16) with a velocity-gradient direction lying within 
45$^\circ$ of the normal to the outflow, and 12 sources out of 14 observed 
interferometrically have the same property (see their Fig.~27). In addition, 
the directions of the single-dish and interferometric gradients are found to
be generally consistent (see their Fig. 28). The average magnitude of the 
single-dish velocity gradients is 2.2~km~s$^{-1}$~pc$^{-1}$, and 8.6 for the
interferometric data. From these results, it is tempting to interprete the 
velocity gradients as due to rotation, with spinning-up from large 
($\sim 10000$~AU) to smaller ($\sim 1000$~AU) scales. However, there is a 
significant spread of position angle relative to the normal to the outflow, 
which may indicate that the velocity gradients do not trace pure rotation. But
if they do, then, again, these offsets imply that the orientation of the 
rotation axis changes from scales of $\sim 10000$~AU to scales of $\lesssim$ a 
few AU where the jets/outflows are launched. The authors suggest also 
that the rotation signature may be affected by the asymmetry of the envelopes, 
most of them being indeed found to be asymmetric \citep[][]{Tobin10a}. Note 
also that both tracers used by \citet{Tobin11} do not peak on the central 
protostar in general, suggesting that they do not trace well the innermost 
parts ($r \lesssim 1000$~AU).

\citet{Chen07} carried out a similar interferometric survey of 8 Class 0 
protostars in N$_2$H$^+$ 1--0. Only two of these sources, which are also in
the sample of \citet{Tobin11}, have a velocity gradient close to the normal to 
the outflow. Combining both samples, only 12 out of 19 Class 0 envelopes
have this property on scales $\sim 2000$--8000~AU. We conclude from this that 
even N$_2$H$^+$ 1--0 is not an ideal probe of rotation in protostellar 
envelopes, 
or, again, that the angular momentum on scales at which the jet/outflow is 
launched ($\lesssim$ a few AU) has a significantly different direction 
than on scales of a few thousand AU. In the latter case, it would be tempting
to conclude that turbulence vorticity, rather than well-ordered rotation, 
still dominates the velocity field on scales of a few thousand AU. It would 
then be interesting to check whether the protostellar envelopes with a velocity 
gradient not perpendicular to the outflow axis are also those with a higher 
level of turbulence.

Finally, in this context, it is worth mentioning that the TADPOL 
polarization survey performed toward a sample of 27 protostellar sources with 
CARMA found a random orientation of the magnetic fields with respect 
to the outflow axes on scales of $\sim 1000$~AU \citep[][]{Hull12}.
This and the results mentioned in the previous paragraphs suggest that the 
naive picture of the rotation, magnetic field, and outflow axes being aligned 
in a protostellar envelope is rare at this scale. For completeness, it would 
be interesting to investigate if there is a degree of correlation between the 
magnetic field direction and the direction of the velocity gradients thought
to trace rotation in these protostellar envelopes.

\subsection{The test case IRAM~04191}
\label{ss:iram04191}

\begin{figure}
%\centerline{\includegraphics[width=0.26\paperwidth,angle=270]{Figs/posvit_mix_mod_art1_ees1209.eps}}
\centerline{\includegraphics[width=0.26\paperwidth,angle=270]{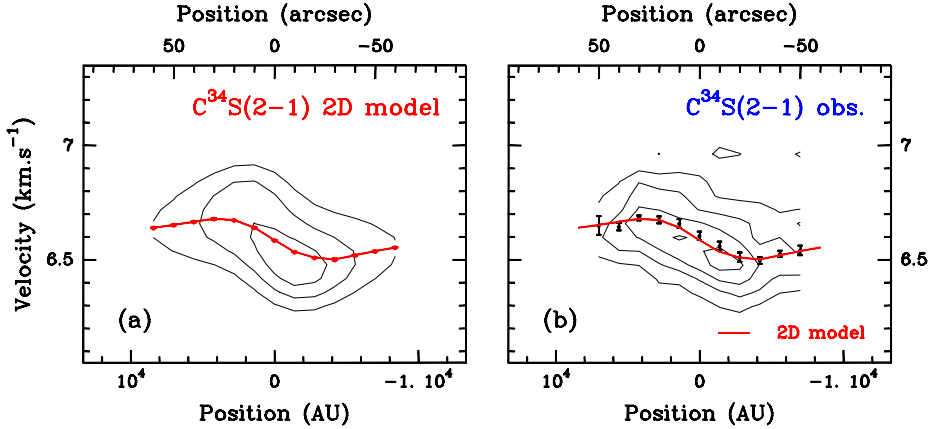}}
\caption{Position-velocity diagram of C$^{34}$S 2--1 along the direction 
perpendicular to the outflow axis in IRAM~04191.
\textbf{a} Synthetic contour map of the ``best-fit'' collapse model with 
differential rotation.
\textbf{b} Observed contour map. The black dots with error bars are the measured
centroid velocities. The red curve in each panel is the synthetic 
centroid-velocity curve. The angular resolution is $25.5''$ ($HPBW$), i.e. 
3600 AU \citep[figure taken from][]{Belloche02a}.}
\label{f:pvdiag_iram04191}
\end{figure}

\begin{figure}
%\centerline{\includegraphics[width=0.50\paperwidth,angle=0]{Figs/fig4-9_th_ab.eps}}
\centerline{\includegraphics[width=0.50\paperwidth,angle=0]{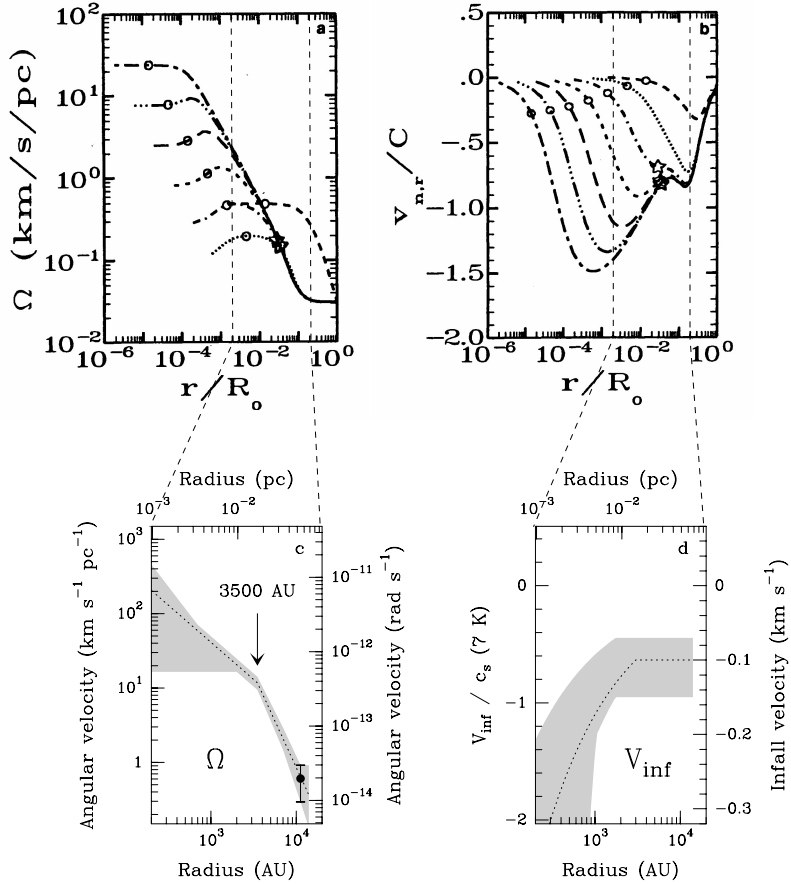}}
\vspace*{-2ex}
\caption{Comparison of the profiles of angular velocity (\textbf{a}) and 
radial velocity (\textbf{b}) of model 8 of \citet{Basu95} with the profiles
(\textbf{c} and \textbf{d}) derived for IRAM~04191
\citep[see caption of Fig. 4-9 of][ for more details]{Belloche02b}.}
\label{f:iram04191_basu}
\end{figure}

IRAM~04191+1522 -- hereafter IRAM~04191 -- is a very young Class 0 protostar 
(1--$3 \times 10^4$~yr) located in Taurus. It has a prominent envelope 
($\sim 1.5\;M_\odot$) and drives a collimated outflow \citep[][]{Andre99}. Its 
internal luminosity is very low \citep[$0.08~L_\odot$,][]{Dunham06}. IRAM~04191 
belongs to the ``well-behaved'' protostellar envelopes in terms of velocity 
gradient: it harbors a regular, large-scale velocity gradient the direction of 
which is parallel to the envelope major axis and orthogonal to the outflow axis 
\citep[see Fig. 2 of][]{Belloche02a}. Its centroid-velocity curve along the
direction of the mean velocity gradient is not a straight line, but it is 
centrosymmetric, which is a strong evidence for rotation (see 
Fig.~\ref{f:pvdiag_iram04191}b). The magnitude of the velocity gradient is 
7--10~km~s$^{-1}$~pc$^{-1}$ at $r \sim 2800$~AU, 1.3~km~s$^{-1}$~pc$^{-1}$ at 
$r \sim 7000$~AU, and $\lesssim 0.8$~km~s$^{-1}$~pc$^{-1}$ at $r \sim 11000$~AU 
\citep[][]{Belloche02a}. The envelope is thus in differential rotation, the 
inner parts rotating faster than the outer parts. The observed P-V diagram is 
well reproduced by a parametric model of a differentially rotating,  infalling,
spherical envelope, with CS (and C$^{34}$S) depleted toward the center 
(see Fig.~\ref{f:pvdiag_iram04191}a). The model fits well several transitions 
of CS and C$^{34}$S and strong constraints on the infall and rotation velocity 
fields could be derived \citep[see Figs.~14 and 12 of][]{Belloche02a}. These 
kinematical features are in qualitative agreement with a model of 
\citet{Basu95} of a supercritical magnetized envelope collapsing and detaching 
from its subcritical environment 
\citep[see][ and Fig.~\ref{f:iram04191_basu}]{Belloche02a}.

The differential pattern seen in the P-V diagram of IRAM~04191 is not unique:
the P-V diagram of CB230 shows a similar shape for instance
\citep[see Fig.~24 of][]{Tobin11}. But only 2--3 protostellar envelopes have
a similar P-V diagram out of 17 observed by \citet{Tobin11}. Therefore the 
conclusions obtained for IRAM~04191 cannot be generalized to all Class 0
protostars.

On smaller scales, the angular velocity of the envelope of IRAM~04191 keeps 
increasing. \citet{Belloche04a} derive a velocity gradient of 
26~km~s$^{-1}$~pc$^{-1}$ at $r \sim 2000$~AU from N$_2$H$^+$ 1--0 
interferometric observations. However, the N$_2$H$^+$ emission map presents a 
hole toward the center and the column density profile is consistent with 
N$_2$H$^+$ being absent from the gas phase for $r \lesssim 1600$~AU 
\citep[see Figs.~1 and 4b of][]{Belloche04b}. Because there is no evidence
for strong C$^{18}$O desorption, \citet{Belloche04b} conclude that N$_2$H$^+$
suffers from depletion at densities higher than $5 \times 10^5$~cm$^{-3}$.
Based on independent N$_2$H$^+$ 1--0 interferometric measurements with similar 
angular resolution, \citet{Lee05} report a linear velocity gradient
for $r < 1400$~AU. They conclude that the innermost parts of the envelope are 
in solid-body rotation, indicating that the infalling material loses angular 
momentum, possibly due to magnetic braking like in the collapse simulations of 
magnetized singular isothermal toroids by \citet{Allen03}. However, an 
alternative (more likely?) explanation is that the depletion of N$_2$H$^+$ 
from the gas phase significantly affects the P-V diagram. If there is no 
molecule emitting below a radius $r_0$, then the P-V diagram for $R < r_0$ 
will trace the velocity field of the shell of radius $r_0$ only, which, in 
projection, will appear as a linear velocity gradient. A tracer less sensitive 
to depletion is thus necessary to unambiguously probe the rotational 
properties of the innermost parts of the IRAM~04191 envelope. H$_2$D$^+$ may 
be a good candidate \citep[][]{Walmsley04}.

\subsection{Do velocity gradients really trace rotation of protostellar envelopes?}
\label{ss:gradv}

\begin{figure}
%\centerline{\includegraphics[width=0.50\paperwidth,angle=270]{Figs/tobin12_fig1.eps}}
\centerline{\includegraphics[width=0.50\paperwidth,angle=270]{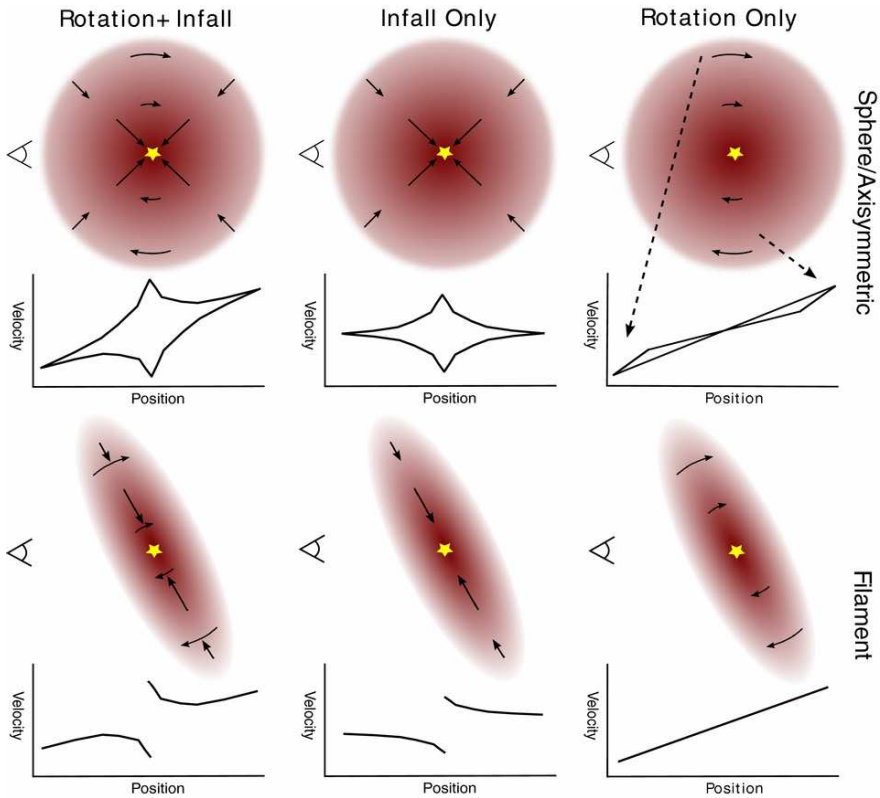}}
\caption{Position velocity diagrams of a spherical or axisymmetric 
envelope (top row) and a filamentary envelope (bottom row), with rotation only 
(right column), infall only (middle column), and rotation and infall 
(left column) \citep[from][]{Tobin12a}.}
\label{f:tobin12}
\end{figure}

The protostellar envelopes studied by \citet{Tobin10a} are
often flattened. They even find indications that some of them are filamentary.
\citet{Tobin12a} argue that infall in such prolate structures at intermediate 
inclination would produce a gradient in centroid-velocity maps that can mimic
(differential) rotation (see Fig.~\ref{f:tobin12}). As a proof of concept, they
use an analytic model of rotating collapse based on \citet{Ulrich76} and 
\citet{Cassen81}, where the self-gravity of infalling gas is neglected. The
particles follow ballistic trajectories around a central gravitating mass.
The model is truncated to mimic a filamentary structure. This model is applied
to 5 sources of \citet{Tobin11} with a well-ordered velocity field. Most
P-V diagrams can be reasonably well fitted with such a model of collapsing 
filament. The agreement between model and observations is not a proof that the
kinematics is effectively dominated by infall in a filamentary structure, but 
it shows that this idea is a viable concept. If envelopes are really 
filamentary, then the velocity gradients could probe infall rather than 
rotation. In this case, it would become necessary to go to smaller scales to 
trace rotation and derive meaningful constraints on the distribution of 
angular momentum.

\subsection{Specific angular momentum in protostellar envelopes}
\label{ss:envelopes_j}

\begin{figure}
%\centerline{\includegraphics[width=0.50\paperwidth,angle=0]{Figs/joverm_ees1209_book.eps}}
\centerline{\includegraphics[width=0.50\paperwidth,angle=0]{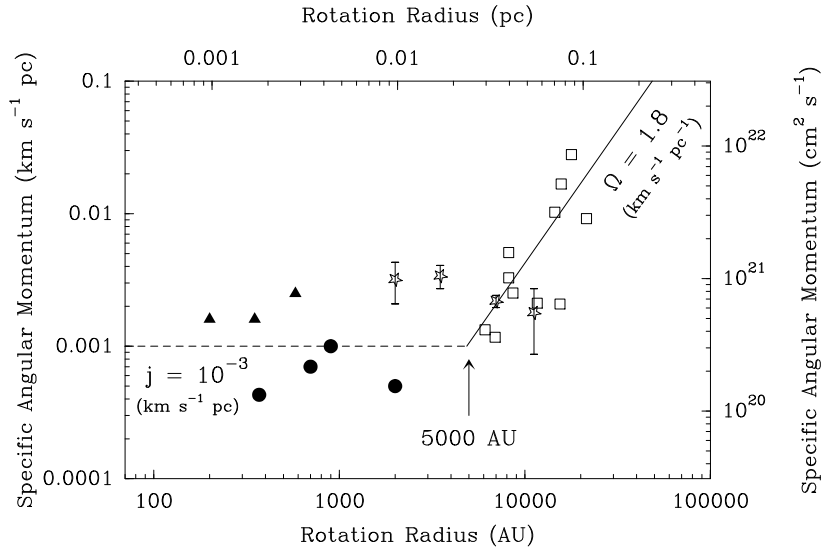}}
\caption{Specific angular momentum as a function of radius for a sample of
sources in Taurus. The squares are NH$_3$ dense cores \citep[][]{Goodman93}. 
The stars are measurements in IRAM~04191 at four different scales 
\citep[][]{Belloche02a,Belloche04a}. The filled circles are rotating, infalling 
envelopes and the triangles are ``rotationally supported disks'', all from 
\citet{Ohashi97} \citep[adapted from][ and \citeauthor{Belloche02a} 
\citeyear{Belloche02a}]{Ohashi97}.}
\label{f:joverm_taurus}
\end{figure}

In this Section, we assume that velocity gradients do trace rotation in 
protostellar envelopes. \citet{Ohashi97} compiled measurements of velocity
gradients in a few protostellar envelopes and circumstellar disks of the Taurus
molecular cloud made in the 90's. The specific angular momenta of these 
systems are compared to those of the NH$_3$ dense cores of \citet{Goodman93}
in Fig.~\ref{f:joverm_taurus}, which can be interpreted as an evolutionary 
diagram. The measurements obtained for IRAM~04191 (see 
Sect.~\ref{ss:iram04191}) are also displayed. The local specific angular 
momentum is approximately constant for infalling envelopes and circumstellar
disks in Taurus (about $10^{-3}$~km~s$^{-1}$~pc or 
$3 \times 10^{20}$~cm$^2$~s$^{-1}$, see dashed line in 
Fig.~\ref{f:joverm_taurus}), while the angular velocity is approximately
constant for the dense core regime (see solid line in
Fig.~\ref{f:joverm_taurus}). The transition between the dense core $j$--$R$
correlation to a regime of constant specific angular momentum occurs at a 
radius of $\sim 5000$~AU or $\sim 0.03$~pc. As pointed out by \citet{Ohashi97},
this appears to be the scale below which the collapse is dynamical with
conservation of angular momentum. The interpretation of the inner envelope of
IRAM~04191 being a magnetically supercritical core detaching from its 
subcritical environment (Sect.~\ref{ss:iram04191}) fits relatively well into 
this pattern\footnote{Beware, however, that the datapoints for IRAM~04191 in
Fig.~\ref{f:joverm_taurus} represent the profile of specific angular momentum 
for an envelope at a given time, while the other points rather correspond to 
the evolution in time of the specific angular momentum of protostellar
systems, which is conceptually different. A direct comparison is therefore 
not straightforward.}. A more complete version of this diagram will be 
presented and discussed in Sect.~\ref{ss:j_evolution}.

\subsection{Summary}
\label{ss:envelopes_summary}

The velocity structure of protostellar envelopes is often complex.
In about $50\%$ of the cases, the velocity gradient in the envelope is not 
orthogonal to the outflow axis. This suggests that either velocity 
gradients do not reliably trace rotation (on scales 1000--10000~AU), or 
the direction of the rotation axis changes from large ($\sim 10000$~AU) to 
small ($\lesssim$ a few AU) scales where the jets/outflows are launched. Given 
that protostellar envelopes are often asymmetric, the velocity gradients may 
also trace infall if the envelopes have a filamentary structure. Therefore, 
velocity gradients do not unambiguously trace rotation in protostellar 
envelopes. However, if the rotation signature in P-V diagrams can be trusted, 
there are good indications for differential rotation, with
spinning-up of the inner parts, to occur below $R \sim 0.05$~pc (e.g., 
IRAM~04191). In addition, there seems to be in Taurus a characteristic scale of 
$R \sim 0.03$~pc below which the specific angular momentum is conserved,
probably as a result of dynamical collapse.

\section{Rotation of protoplanetary disks}
\label{s:disks}

\subsection{General properties of circumstellar disks}
\label{ss:disks_properties}

The presence of disks aroung young pre-main-sequence (PMS) stars was initially 
inferred from the infrared excess emission in the spectral energy 
distribution of these stars. The flattened morphology of these putative disks 
was later confirmed by imaging, in particular as shadows in front of a bright 
background in the optical \citep[see, e.g., Figs.~1 of][ and 
\citeauthor{Smith05} \citeyear{Smith05}]{McCaughrean98}. The 
lifetime of disks around solar-type PMS stars 
is on the order of 1--10~Myr with a median of 3~Myr \citep[][]{Williams11}. 
Their mass ranges from 0.5 to 50~$M_{\rm jup}$ with a median of 
5~$M_{\rm jup} = 0.005~M_\odot$ \citep[][]{Williams11}. Given the high detection 
rate of exoplanets around stars \citep[e.g.,][]{Mayor11}, it is reasonable to
consider the circumstellar disks around PMS stars as protoplanetary.

In the Class~II phase, the median ratio of disk to star masses is 0.9\%
\citep[][]{Williams11}. Therefore the disk self-gravity is negligible and we
expect their kinematics to be dominated by Keplerian motion, i.e. 
$v_{\rm rot} = \sqrt{\frac{GM_\star}{r}}$. Centroid-velocity maps of such disks
are expected to have a very characteristic bipolar morphology 
\citep[see, e.g., Fig.~2 of][]{Guilloteau06}. Since Keplerian disks exist 
in the Class~II phase (see Sect.~\ref{ss:disks_class2}), they must have formed 
during the Class 0/I protostellar phase. The next three sections will try to 
address the questions of when a disk appears in the process of star formation 
and when Keplerian rotation sets in.

\subsection{Disk formation}
\label{ss:disks_formation}

The centrifugal radius is defined as the radius where gravity is balanced by
the centrifugal force $\frac{v^2_{\rm rot}}{R}$. For a collapsing, non-magnetized,
singular isothermal sphere initially in solid-body rotation, the centrifugal
radius is found to grow with time as:
\begin{equation}
R_{\rm c}({\rm AU}) \sim 39 \left(\frac{\Omega}{10^{-14}\,{\rm rad~s}^{-1}}\right)^2 \left(\frac{a}{0.2\,{\rm km~s}^{-1}}\right)^{-8} \left(\frac{m_{\star+\rm d}}{1~M_\odot}\right)^3\,,
\end{equation}
with $m_{\star+\rm d} = 0.975 \frac{a^3}{G} t$ and $a$ the sound speed 
\citep[][]{Terebey84}. Under these restrictive and idealized conditions, 
rotationally supported disks are expected to form from the beginning of the 
Class~0 phase on with a radius growing as $t^3$. As an illustrative 
application, let's compute the expected 
centrifugal radius of IRAM~04191 in the framework of this model. We use the 
velocity gradient measured at 2000~AU to estimate $\Omega$ (see 
Sect.~\ref{ss:iram04191}). For a temperature of 10~K and an age of 
1--$3 \times 10^4$~yr, we find $R_{\rm c} = 3$--80~AU. 
%(26/math.sin(50./180.*math.pi)*1.e3/3.0857e16/1.e-14)**2*(0.975*200**3/6.67e-11*3e4*86400*365.25/1.989e30)**3*39
The constraints derived on the kinematics of the envelope imply an upper limit 
to the centrifugal radius of 400~AU \citep[see Fig.~15 of][]{Belloche02a}.
In addition, the weak interferometric detection in continuum emission at
1.3~mm implies either a disk mass lower than 0.001~$M_\odot$ or a disk radius 
smaller than 10~AU if the emission is optically thin or thick, respectively 
\citep[][]{Belloche02a}. Depending on the exact age of IRAM~04191, the latter
upper limit is marginally consistent with the prediction of \citet{Terebey84}.
 
The non-magnetized picture of disk growth described in the previous paragraph
has been challenged in the past decade by the advent of 3D MHD simulations.
In ideal MHD, for magnetization levels comparable to those observed in dense
cores, these simulations predict a ``magnetic braking catastrophe''
which prevents the formation of a disk when the magnetic field is parallel to 
the rotation axis \citep[e.g.,][; see also the axisymmetric simulations of 
\citeauthor{Allen03} \citeyear{Allen03}]{Mellon08,Hennebelle08}. 
When the initial magnetic and rotation axes are not aligned, disk formation 
can occur depending on the initial mass-to-magnetic-flux ratio 
\citep[][]{Hennebelle09,Joos12}. Non-ideal MHD effects such as ambipolar 
diffusion, Ohmic dissipation, and the Hall effect, have also been investigated 
recently
\citep[e.g.,][]{Shu06,Mellon09,Krasnopolsky10,Dapp10,Machida11,Krasnopolsky11,Li11}.
Ambipolar diffusion does not change the disk outcome, but Ohmic dissipation and
the Hall effect may (or not) enable the formation of rotationally supported 
disks in the Class~0 phase, albeit maybe small ones only 
\citep[$<$ a few AU, see discussion in][]{Li11}. This is currently an active 
field of research and the conclusions are not well settled. Observational
constraints on the formation of disks in the Class~0 phase are highly needed.

\subsection{Class 0 protostellar disks (?)}
\label{ss:disks_class0}

Several interferometric surveys were performed in dust continuum emission to 
search for disks around Class~0 and I protostars. The PROSAC survey was 
performed at 1.1~mm with the SMA with an angular resolution of 1--3$''$ (i.e. 
not sufficient to resolve most disks), with complementary single-dish maps 
obtained at 870~$\mu$m with SCUBA on JCMT. The sample includes 10 Class~0 and 
9 Class~I protostars at distances ranging from 125 to 325~pc 
\citep[][]{Jorgensen09}. The respective contributions of the envelope and the 
(unresolved) ``disk'' to the continuum emission was estimated using a 
parametric model of spherically-symmetric protostellar envelopes with a 
power-law density structure. Assuming 30~K for the disk emission and the 
same dust opacity coefficient as for the envelope, \citet{Jorgensen09} derived 
a median disk mass of 0.09~M$_\odot$ for Class~0 and 0.011~$M_\odot$ for Class~I
objects. After corrections for non-uniform temperature, they obtain the same
median disk mass for Class~0 and I protostars ($\sim 0.04~M_\odot$). Taking 
grain growth into account would lower the Class~I median mass with respect to
the Class~0 one. It appears from this study that (massive) disks are formed
early in the Class~0 phase and that the infalling material from the 
envelope must be rapidly transported through the disk onto the star.

\citet{Enoch11} surveyed 9 Class~0 and 3 Class~I protostars in Serpens with 
CARMA at 1.3~mm with an angular resolution of $\sim 1''$. They estimated the
``disk'' masses from the long-baseline (50~k$\lambda$) flux, arguing that the 
envelope emission contributes to at most 30\% and can be neglected. They
detected 6 Class~0 sources and obtained a median disk mass of 0.15~$M_\odot$.
Based on this small sample, they did not see any obvious systematic variation 
of the disk mass from the Class~0 to the Class~I phase, in agreement with 
\citet{Jorgensen09}.

\citet{Maury10} compare the results of their sub-arcsecond resolution 
\hbox{($HPBW = 0.3$--0.5$''$)}, dust continuum  survey of 5 Class~0 protostars 
in Taurus and Perseus performed with the IRAM Plateau de Bure interferometer
(PdBI) to 
the outcome of hydrodynamic simulations of a massive (0.7~$M_\odot$) disk around
a 0.7~$M_\odot$ star, which fragments rapidly and may thus exist only during the
Class~0 phase \citep[][]{Stamatellos09}. They conclude that the data are not 
consistent with the model and that none of their sources can harbor 
such a massive disk. However, they do not exclude the presence of 
0.1~$M_\odot$ disks.

Although the results of \citet{Jorgensen09} and \citet{Enoch11} tend to 
suggest that $\sim 0.1~M_\odot$ ``disks'' are common in the Class~0 phase, a 
significant number of Class~0 protostars do not show any evidence for the 
presence of such disks: $M_{\rm disk} < 0.02~M_\odot$ 
for 3 sources in Serpens \citep[][]{Enoch11}, $< 0.01~M_\odot$ for L483 
\citep[][]{Jorgensen09}, $< 0.006~M_\odot$ for L723 \citep[][]{Girart09}, and 
$< 0.001~M_\odot$ for IRAM~04191 \citep[][]{Belloche02a}. L1521F is even weaker
than IRAM~04191 in the survey of \citet{Maury10}, suggesting that its disk,
if any, is even less massive.

Although \citet{Jorgensen09} and \citet{Enoch11} claim detecting 
$\sim 0.1~M_\odot$ ``disks'' in their samples of Class~0 sources, it is 
important to keep in mind that the emission was not resolved by these 
observations and that strong assumptions about the structure of the envelope 
-- spherically-symmetric envelope with power-law structure for the former 
study, no envelope 
contribution at 50~k$\lambda$ for the latter -- had to be made in order to 
derive the excess emission attributed to a ``disk''. \citet{Chiang08} compared
the emission of 4 Class~0 sources observed by \citet{Looney00} at 
sub-arcsecond resolution ($HPBW \sim 0.6''$) with BIMA -- 3 being also in the 
PROSAC sample -- with a model of a collapsing, magnetized envelope 
\citep[][]{Tassis05}. They obtain good fits for all their sources without the 
need for an additional circumstellar-disk component at the 90\% confidence 
level, which contradicts the conclusions of \citet{Jorgensen09} for the 
sources that both studies have in common. This shows that knowing the inner 
structure of the envelope, which departs from a power-law in the model of
\citet{Tassis05}, is crucial to derive reliable constraints on the mass of a
putative disk. \citet{Chiang08} mention that models with a disk mass up to 
$\sim 0.1~M_\odot$ in addition to the magnetized envelope give acceptable fits 
also.

The  unresolved components interpreted as disks by \citet{Jorgensen09} and 
\citet{Enoch11} may actually be infalling, magnetized pseudo-disks 
\citep[e.g.,][]{Hennebelle08}, i.e. structures that become flattened because of 
the magnetic field rather than rotation. The observed unresolved 
components could also be related to outflow cavities rather than circumstellar 
disks \citep[][]{Joos12}. Higher-angular resolution is necessary to better
resolve the continuum emission and allow for a detailed comparison to
predictions of collapse models. In addition, probing the velocity field of 
these compact structures will be crucial to distinguish between the
alternatives mentioned above: for instance, a magnetic pseudo-disk should be 
\textit{infalling} while the velocity field of a rotationally supported disk 
should be dominated by rotation.

One of the highest-resolution study of the kinematics of the inner parts of 
a Class~0 protostellar envelope was done by \citet{Choi10}. They mapped the 
Class~0 binary system NGC~1333~IRAS~4A in NH$_3$ (2,2) and (3,3) with VLA with 
an angular resolution of 0.3$''$, i.e. 70~AU at the distance of 235~pc which 
they assume for Perseus. They detect a compact emission of deconvolved size 
$130 \times 70$ AU around the component 4A2 and find a clear velocity 
gradient perpendicular to the outflow axis (see their Fig.~2). They use a 
parametric model of a thin disk assuming a power-law distribution for the 
NH$_3$ line brightness to fit the velocity field of this compact structure. 
They reproduce well the position-velocity diagram taken along the direction 
perpendicular to the outflow \textit{assuming} a rotation velocity 
$v_{\rm rot} \propto r^{-0.5}$. Assuming the structure is in Keplerian rotation,
they derive the mass of the central star ($0.08 \pm 0.02~M_\odot$) and then
its age ($5 \times 10^4$~yr) from an estimate of the mass accretion rate
($2 \times 10^{-6}\,M_\odot$~yr$^{-1}$) based on the measured accretion 
luminosity and assuming $R_\star = 2~R_\odot$. This age is somewhat longer than 
the dynamical time associated with the outflow 
\citep[$1.5 \times 10^4$~yr,][]{Knee00}, but is not inconsistent given the large
uncertainties on the luminosity and the stellar radius. This could be 
the first reliable observational evidence for a rotationally supported disk in
the Class~0 phase. We note however that the mass infall rate measured in
the collapsing envelope is about 50 times higher than the mass accretion rate 
mentioned above \citep[][]{DiFrancesco01,Belloche06}. The mass accretion rate 
may thus have been underestimated and radial motions in the ``disk'' may still 
be significant, implying in turn that the measured rotation velocities may be 
sub-Keplerian and not Keplerian. In any case, a better resolved velocity 
profile is needed in order to \textit{fit} the exponent of the rotation 
velocity profile.

Very recently, \citet{Tobin12b} reported the detection of a Keplerian disk
around the Class 0 protostar L1527 based on SMA and CARMA observations. This
system is known to be edge-on with infrared images revealing on very small 
scales a 60 AU thick dark lane that suggests a disk seen in absorption against 
scattered light from the central protostar. The SMA continuum observations with 
$0.25''$ angular resolution resolve a flattened structure coinciding with the 
infrared dark lane, with a radius of 90~AU and a mass of 
$0.007 \pm 0.0007$~$M_\odot$. The $^{13}$CO 2--1 CARMA observations show a clear
velocity gradient along the flattened structure. The angular resolution is 
about $1''$ (140~AU), but the location of the emission peak in each velocity
channel was determined with higher precision, which allowed the authors to 
derive a velocity curve consistent with $v_{\rm rot} \propto r^{-0.5}$. Assuming 
the disk is rotationally supported, they derive a mass of 
$0.19 \pm 0.04$~$M_\odot$ for the central protostar, and then an age of 
$3 \times 10^5$~yr from an estimate of the mass accretion rate 
($6.6 \times 10^{-7}\,M_\odot$~yr$^{-1}$) based on the measured accretion 
luminosity and assuming $R_\star = 1.7$~$R_\odot$. This age is likely to be 
overestimated because it assumes a constant accretion rate. The dynamical
timescale associated with the outflow is indeed an order of magnitude shorter. 
Like for NGC~1333~IRAS~4A, higher-angular-resolution molecular-line 
observations will be decisive to \textit{fit} the exponent of the velocity 
profile and fully demonstrate the Keplerian nature of this disk. L1527 has the
advantage of being a factor of $\sim 2$ closer.

\subsection{Class I protostellar disks}
\label{ss:disks_class1}

As was seen in Sect.~\ref{ss:disks_class0}, the detection of rotationally
supported disks in the Class~0 phase is challenging because the envelope is 
still prominent at this stage and its structure needs to be well understood in 
order to disentangle the contribution of a disk to the detected emission. From 
this point of view, the Class~I phase is \textit{a priori} more favorable 
because the mass of the central object (star plus disk) now dominates the
system.

The kinematics of the young ($T_{\rm bol} = 95$~K) and luminous 
($L_{\rm bol} = 33~L_\odot$) Class~I protostar L1551-IRS5 in Taurus was studied 
in C$^{18}$O 1--0 with the Nobeyama Millimeter Array (NMA)
($HPBW$ $2.8'' \times 2.5''$) by \citet{Momose98}. The continuum emission
is elongated roughly perpendicular to the outflow and a clear velocity gradient
is detected (see their Figs.~1 and 3). Its direction makes an intermediate 
angle to the outflow axis, which is interpreted as resulting from a 
combination of infall and rotation. \citet{Momose98} interprete the 
position-velocity diagrams in terms of a flattened envelope inclined to the 
line of sight.
Infall produces a velocity gradient along the minor axis of the continuum 
emission and rotation along the major axis. They derive an infall velocity 
$v_{\rm inf} \propto r^{-0.5}$. The rotation velocity is found to be 
$v_{\rm rot} \propto r^{-1}$ in the outer parts ($r > 5''$) and consistent with 
both $r^{-1}$ and $r^{-0.5}$ in the inner parts. The authors estimate that 
contamination by the outflowing gas is negligible and derive a mass infall 
rate of $6 \times 10^{-6}~M_\odot$~yr$^{-1}$. They estimate the centrifugal 
radius to be $\sim 160$~AU, the radius were the infall and rotation velocities 
are approximately equal. A similar analysis was done for a small C$^{18}$O 1--0
survey of 2 Class~0 and 6 Class~I protostars in Taurus performed with
the NMA and OVRO \citep[][]{Ohashi97,Ohashi99}. The infall and rotation
velocity profiles were derived from P-V diagrams along the
minor and major axes, respectively. They found 3 sources with both infall
and rotation signatures. From an extrapolation of the derived velocity
profiles, they find centrifugal radii in the range 100--170~AU. These
radii may characterize the rotationally-supported disks in the Class~I phase
but it should be kept in mind that they result from an extrapolation and are
therefore relatively uncertain. It is in any case not a direct proof of the
existence of such disks.

The Class~I sources included in the PROSAC sample (see 
Sect.~\ref{ss:disks_class0}) were also observed with 
SMA in HCO$^+$ 3--2 at a resolution of $\sim 3''$, i.e. about 400~AU 
\citep[][]{Brinch07,Lommen08,Jorgensen09}. The HCO$^+$ emission was found to
be elongated like the continuum emission for 4 out of 10 sources. For the
other sources, it may be contaminated by the outflow. The P-V diagrams along
the direction perpendicular to the outflow were compared to Keplerian
velocity profiles for these 4 sources. The authors find such velocity 
profiles to be consistent with the data, suggesting that they may trace
rotationally-supported disks (of sizes $\sim 500$~AU). However, these 
comparisons are not a proof that the ``fit'' is unique. Other velocity 
profiles may also be in reasonable agreement. There is a clear need for 
better resolved rotation velocity profiles for all Class~I sources discussed 
up to here.

HH~111 is a young ($T_{\rm bol} = 78$~K) Class~I system driving a highly 
collimated jet in L1617 in Orion \citep[see Fig.~1 of][]{Reipurth99}. 
\citet{Lee11} resolved the inner continuum emission as a flattened structure
(deconvolved $FWHM$ $240 \times 120$~AU) perpendicular to the jet. The envelope
is rotating, as traced in C$^{18}$O 2--1 emission with SMA 
\citep[$HPBW \sim 1.2''$, i.e. 500~AU,][]{Lee10}. The P-V diagram along the 
direction perpendicular to the jet shows a clear velocity pattern with
$v_{\rm rot} \propto R^{-1}$ for $R > 2000$~AU and $v_{\rm rot} \propto R^{-0.5}$
for $R < 2000$~AU (see their Figs.~4 and 5). This P-V diagram and the one along
the outflow axis are fitted with an LTE radiative-transfer model, which 
provides a good agreement, apart from some contamination by the outflow in the
latter (see their Fig.~7). The specific angular momentum is found to be 
uniform ($8 \,\, 10^{-3}$~km~s$^{-1}$~pc$^{-1}$) over the range 2000--8000~AU,
and decreasing toward the center at smaller radii. At higher-angular 
resolution with SMA ($HPVW = 0.6''$, i.e. 240~AU), $^{13}$CO 2--1 is, apart
from some outflow contamination, well fitted by the rotation velocity
profile derived above from C$^{18}$O and suggests the presence of a resolved
Keplerian disk \citep[][]{Lee11}. From the continuum emission the author 
estimates a disk mass of 0.14~$M_\odot$ and from the rotation velocity profile 
($\propto r^{-0.5}$) in the inner part a stellar mass of 1.3~$M_\odot$. In 
addition, \citet{Lee10} argues that the compact emission probed with SO may 
trace an accretion shock at 400~AU. The picture emerging from these results is 
that HH~111 has an outer (2000--8000~AU) envelope collapsing with conservation 
of specific angular momentum, a transitional, sub-Keplerian, collapsing inner 
(400--2000~AU) envelope, and a rotationally-supported disk of radius 400~AU.
One puzzle in this picture is that the rotation velocities are found to 
dominate over the infall velocities up to at least 8000~AU in the envelope.

Finally, let us discuss the structure of L1551-NE, a young 
($T_{\rm bol} = 91$~K) Class~I binary system in Taurus. \citet{Takakuwa12}
observed it in 0.9~mm continuum emission and in C$^{18}$O 3--2 with SMA at
very high angular resolution ($HPBW \sim 0.7''$, i.e. 100~AU). The continuum
emission is ring-like, with a strong peak in the center, and is interpreted 
as consisting of (unresolved) circumstellar disks and a (resolved) circumbinary
disk. The C$^{18}$O emission is modeled with both a thin Keplerian or infalling
disk. The best-fit is obtained for a Keplerian-disk velocity structure without
infall. From this modeling, the authors derive a circumbinary disk radius of
300~AU, a total (circumbinary+circumstellar) disk mass of 0.05~$M_\odot$ and
a stellar (binary) mass of $0.8~M_\odot$.

The presence of rotationally supported disks is proven rather convincingly  
for the last two sources, HH~111 and L1551~NE, with radii of 300--400~AU. Both 
sources have a bolometric temperature close to the threshold separating
Class~I from Class~0 sources (70~K, but with the caveat that $T_{\rm bol}$ is
sensitive to the inclination). The presence of extended Keplerian disks around 
such very young Class I sources strongly suggests that these extended disks 
were formed during the Class 0 phase already. This holds true only if these 
sources are as young as suggested by their bolometric temperatures. The PROSAC
sample contains good disk candidates, but higher-angular resolution 
observations of optically thin tracers are needed to confirm the Keplerian
rotation in these objects. Overall, the studies mentioned in this section
suggest Class~I disk radii in the range 100--400~AU and a disk-to-star mass
ratio in the range 1--10$\%$, somewhat higher than for Class~II sources 
(median ratio 0.9\%, see Sect.~\ref{ss:disks_properties}).

\subsection{Class II protoplanetary disks}
\label{ss:disks_class2}

While Keplerian rotation in circumstellar disks is still laborious to probe in 
the Class~0 and I phases, it is observationally well established around 
T~Tauri stars. This is for instance well shown by the CO 2--1 survey of 
T~Tauri disks in Taurus carried out with the PdBI at 
an angular resolution of $\sim 0.7''$ (100~AU) by \citet{Simon00}. They 
performed a $\chi^2$ minimization of a parametric model of a disk in
hydrostatic equilibrium \citep[][]{Dutrey94,Guilloteau98}. The velocity field 
is parametrized as $v_{\rm rot} \propto r^{-\alpha}$ and the best fit for all 
sources is obtained for $\alpha = 0.5$, with a precision better than 10\%. 
This allows for a precise measurement of the stellar masses.

Three recent surveys of PMS circumstellar disks were made in continuum emission:
14 low- and intermediate mass PMS stars (10 in Taurus) with CARMA
\citep[$HPBW = 0.7''$, i.e. 100~AU,][]{Isella09}, 17 PMS stars in Ophiuchus
with SMA \citep[$HPBW = 0.3''$, i.e. 40~AU,][]{Andrews09, Andrews10}, and
23 PMS stars in Taurus-Auriga with the PdBI at two 
frequencies \citep[$HPBW$ down to $0.4''$, i.e. 60~AU,][]{Guilloteau11}. 
For the first two datasets, the surface density profile was derived by fitting a
parametric model of a 2D, flared density structure based on a similarity
solution describing the viscous evolution of an accretion disk, with a
viscosity scaling as $\nu \propto R^\gamma$ \citep[][]{LyndenBell74,Hartmann98}.
The radiative transfer was performed with a two-layer approximation for the
first study and in full 2D for the second study. The dust opacity coefficient
$\kappa$ and the gas-to-dust mass ratio were assumed to be uniform. The
best-fit models of the first study yield disk radii in the range 90--320~AU, 
initial radii $R_1 \sim 25$--40~AU (63\% of the mass contained within $R_1$,
90\% within $2 R_1$), and initial masses in the range 0.05--0.4~$M_\odot$.
Making a number of assumptions, \citet{Isella09} estimate the specific
angular momenta required for parent dense cores to produce disks with initial
radii within the range they derive. They find 
$j_{\rm core} \sim 0.8$--$4 \,\, 10^{-4}$~km~s$^{-1}$~pc, 10 times smaller than
the observed specific angular momenta of N$_2$H$^+$ dense cores of mass 
1--10~$M_\odot$. They conclude that only 10\% of the core specific angular 
momentum was conserved during the collapse phase. Following a similar procedure 
with slightly different assumptions, \citet{Andrews10} derive 
$j \lesssim 1.6$--$26 \,\, 10^{-4}$~km~s$^{-1}$~pc for their Ophiuchus sample.
They argue that this matches the lower range of N$_2$H$^+$/NH$_3$ dense cores,
and that the agreement may even be better if the observed dense-core 
$j_{\rm 2D}$ overestimate the true $j_{\rm 3D}$ (see 
Sect.~\ref{ss:j2d_reliability}). It
is therefore unclear whether these disks measurements imply a loss of angular 
momentum during the collapse phase or not.

The PdBI sample was fitted with two models, a truncated power-law model and
a viscous disk model similar to the previous two studies. \citet{Guilloteau11}
find that both models fit the data equally well. The measurements were done at
two frequencies, allowing the dust opacity index $\beta$ 
($\kappa \propto \nu^\beta$) to be determined across the disks. $\beta$ is 
found to increase from $\sim 0$ at center to 1.7--2 at the edges, suggesting
that grains are larger toward the center. The disk outer radii are found in the
range 14--600~AU, with a hint of increase with stellar age, which would
be consistent with a viscous evolution. Based on the viscous disk model,
the authors derive initial disk radii smaller than 100~AU.

All three studies find initial and current outer radii of PMS disks somewhat 
smaller than the disk radii derived for the Class~I objects 
($\sim$ 100--400~AU, see Sect.~\ref{ss:disks_class1}). This may suggest that
the Class~I ''disks'' are not yet fully rotationally supported disks or that
they are pseudo-disks (see definition in Sect.~\ref{ss:disks_class0}). 
Alternatively, inferring disk evolution from current surface
density profiles of observed PMS disks based on a viscous disk model may not
be correct. Finally, the samples are still small, especially for the Class~I
phase, and may not be free of biases.

\subsection{Summary}
\label{ss:disks_summary}

The theoretical outcome of dense core collapse is uncertain in terms of
disk formation: the existence of disks and their properties depend strongly
on the magnetic field configuration and on the role played by non-ideal MHD
effects. The existence of rotationally supported disks in the Class 0 phase
is not fully established observationally, even if two sources turn out to show
good Keplerian-disk candidates. Higher angular resolution is needed to 
distinguish Keplerian disks from pseudo-disks or envelope small-scale 
structures. 
''Disks'' associated with Class~I protostars have typical radii of 100--400~AU
and disk-to-star mass ratios of 1--10$\%$. The presence of a rotation velocity
field scaling as $r^{-0.5}$ is well demonstrated in a few cases, possibly
very young Class~I objects, suggesting that these disks were already formed
during the Class~0 phase. The scaling as $r^{-0.5}$ may however not be an
unambiguous sign of \textit{Keplerian} rotation. 
Keplerian rotation is well established in disks around PMS stars. The small
radii of these PMS disks question the Class~I disks being already Keplerian.
The latter may correspond to pseudo-disks.

\section{Rotation of jets}
\label{s:jets}

\subsection{Properties and origin of jets}
\label{ss:jets_properties}

Jets and outflows are ubiquitous in star formation. Most YSOs where accretion
or infall is occuring are associated with a jet or an outflow 
\citep[][]{Cabrit02}: Class~0 and I protostars, which have strong or residual 
infall in their envelope plus possibly an accretion disk, and Class~II PMS 
stars, which have accretion disks. A strong correlation between ejection and
accretion was established in Class~II sources, suggesting that the ejection 
process is driven by disk accretion \citep[][]{Cabrit90b}. The ratio
$\frac{2 \dot{M}_{\rm ej}}{\dot{M}_{\rm acc}}$ is about 0.1--0.2 in Class~II 
sources \citep[][]{Cabrit07}. Jets from Class~II sources have narrow opening 
angles (a few degree) beyond 50~AU and the sources have no dense envelopes to 
confine them, which means that they must be intrinsically collimated. It is 
believed that this happens through MHD self-collimation along rotating open 
field lines \citep[e.g.,][]{Ferreira06}. Jets are thought to be driven by 
magnetocentrifugal acceleration, 
which implies that they remove angular momentum from the accretion disk and/or 
the stellar magnetosphere \citep[see, e.g.,][, and also the contribution of
J. Ferreira in this volume]{Konigl00,Shu00, Zanni13}.

Three main magnetocentrifugal mechanisms have been proposed for steady jets
\citep[see, e.g.,][ for details]{Konigl00,Shu00,Ferreira06}: the stellar wind
model which launches the jet close to the stellar surface, the X-wind 
model for which the jet is launched from a narrow region close to the radius 
$R_{\rm X}$ of the disk where it is truncated by the stellar magnetosphere, 
and the extended disk wind model for which the launching occurs over a wider 
range of disk radii ($> R_{\rm X}$). Constraints on the launching mechanism, 
such as the radius
where it occurs and the magnetic lever arm braking the rotating disk/star, can
be derived from the location of the jet in a diagram displaying the local 
specific angular momentum of the jet ($Rv_\phi(R)$, with $v_\phi(R)$ the 
toroidal velocity) versus its poloidal velocity
($v_p(R)$) \citep[see Fig.~2 of][]{Ferreira06}. This motivates the search for
rotation signatures in jets. 

\citet{Anderson03} computed a practical form of the equation 
relating the launching radius $R_0$ to the poloidal and toroidal velocities of
the jet, $v_p(R)$ and $v_\phi(R)$, at a distance $R$ from its axis and far 
from the launching region (their Equation 5): 
\begin{equation}
\label{eq:r0}
R_0 \approx 0.7\;{\rm AU} \left( \frac{R}{10\;{\rm AU}} \right)^{2/3} 
\left( \frac{v_\phi(R)}{10\;{\rm km~s}^{-1}} \right)^{2/3}
\left( \frac{v_p(R)}{100\;{\rm km~s}^{-1}} \right)^{-4/3} 
\left( \frac{M_\star}{1\;M_\odot} \right)^{1/3} .
\end{equation}
As stated by \citet{Anderson03}, this equation is an approximation and is 
valid only when the kinetic
energy of the jet is much greater than the gravitational binding energy at
the launching region. It is not valid anymore when the magnetic lever arm is
small, in which case more terms have to be taken into account (see their
Equation 4). This is the reason why the curves of constant $R_0$ in Fig.~2 of
\citet{Ferreira06} are curved toward low values of the magnetic lever arm 
($\lambda$) while Equation~\ref{eq:r0} would produce straight lines. After
using Equation~\ref{eq:r0}, it should thus always be verified a posteriori that
the gravitational energy at the launching region is negligible compared to the 
kinetic energy of the jet, otherwise the complete equation has to be used.

\subsection{Jet rotation signature}
\label{ss:jets_rotsignature}

In the absence of rotation, spectra taken at symmetric position on each side
of the axis of an axisymmetric jet should have the same centroid velocity.
If the jet is rotating, there should be a velocity shift between both spectra.
\citet{Pesenti04} present predictions for an optical line ([O I] at 
$\lambda6300$) based on a self-similar MHD disk-wind solution and taking into 
account projection effects (inclination angle $i$) and a finite angular 
resolution. They investigate two ionization profiles, 
$x_{\rm e} \propto \frac{1}{R_0}$ and $x_{\rm e} \propto R_0$, $R_0$ being the 
radius in the disk. Let's call $v_R$ 
and $v_{-R}$ the centroid velocities of spectra taken at symmetric positions
at distances $+$/$-R$ on each side of the jet axis, $v_{\rm p}(R)$ and 
$v_{\rm \phi}(R)$ the poloidal and azimuthal velocities of the jet, and 
$v_{\rm shift}(R)$ the velocity offset between 
both spectra. Under the assumption that only one streamline dominates the 
emission along each line of sight, we have the following equations:
\begin{equation}
\label{eq:vp}
v_{\rm p}(R) = \frac{1}{\cos{i}} \frac{v_R+v_{-R}}{2}\;,
\end{equation}
\begin{equation}
v_{\rm \phi}(R) = \frac{1}{\sin{i}} \frac{v_R-v_{-R}}{2} = \frac{1}{\sin{i}} \frac{v_{\rm shift}(R)}{2}\;.
\end{equation}
Equation~\ref{eq:vp} assumes that $v_{\rm p} \sim v_{\rm z}$, i.e. that the 
radial velocity $v_{\rm r} << v_{\rm p}$.
As shown in Fig.~3b of \citet{Pesenti04}, $v_\phi(R)$ can be underestimated if 
the jet section is not well resolved (for $R \lesssim 2\,R_{beam} = HPBW$). 
This implies that only the external radius of the launching region can often
be reliably determined.
$v_\phi(R)$ is also sensitive to the 
distribution of the line emission across the jet axis because in reality 
several streamlines contribute along each line of sight. If the outer, slower
streamlines dominate the emission, then $v_\phi(R)$ will also  be underestimated
\citep[see the difference between the predictions for 
$x_{\rm e} \propto \frac{1}{R_0}$ and $x_{\rm e} \propto R_0$ in Fig.~3b 
of][]{Pesenti04}. As a result, a careful choise of the tracer and a high 
angular resolution (better than 5~AU, i.e. $0.04''$ at 125~pc) are needed to
fully probe the rotation \textit{profile} of the jet and test MHD launching 
models.

\subsection{Jet rotation in the Class II and I phases}
\label{ss:jets_class1and2}

Indications of jet rotation based on measurements of transverse velocity 
shifts were found in a few Class I and II sources (see 
Sect.~\ref{ss:jets_statistics}).
One of the first claimed evidences for rotation in a jet was presented in 
\citet{Bacciotti02} for the Class~II object DG~Tau 
\citep[see][ for an earlier claim in HH~212]{Davis00}. The authors observed 
with a slit parallel to the jet, at seven offsets across the jet, in optical 
forbidden lines ([O~I], [N~II], and [S~II]) with an angular resolution of 
$\sim 0.1''$, i.e. about 14~AU. They applied two methods, multiple Gaussian
fitting and cross-correlation of spectra at symmetric positions with respect
to the jet axis, both with an accuracy of $\sim 5$~km~s$^{-1}$ (see their 
Fig.~1). They measure velocity shifts 
$v_{\rm shift} \sim 5$--20~$\pm 5$~km~s$^{-1}$ across the jet (see their Fig.~2) 
and conclude that the jet is rotating with $v_\phi \sim 6$--15~km~s$^{-1}$ for 
distances from the central star between $0.075''$ and $0.4''$, i.e. 11 and 
56~AU (in projection). In the framework of the jet launching paradigm 
described in Sect.~\ref{ss:jets_properties}, the radius of the launching 
region is $R_0 \sim 3$~AU for the external part of the atomic jet \citep[][, 
with a more sophisticated analysis than in \citeauthor{Bacciotti02} 
\citeyear{Bacciotti02} and \citeauthor{Anderson03} 
\citeyear{Anderson03}]{Pesenti04}. In addition, 
\citet{Bacciotti02} estimate that the flux of angular momentum carried 
by the jet represents about 60\% of the angular momentum loss rate needed in 
the disk at $R_0$ for accretion to occur at the observed rate. Further 
observations with HST/STIS in the near-UV confirmed the sign and magnitude of 
the velocity gradient in DG~Tau \citep[][]{Coffey07}. 
Signatures of rotation were also found for the molecular (H$_2$) jet of the 
Class~I protostar HH~26 with observations performed in H$_2$ 1--0 S(1) with 
VLT/ISAAC \citep[][]{Chrysostomou08}. The authors derive $R_0 \sim 2$--4~AU 
and estimate that the transported angular momentum flux represents about 70\%
of the loss rate needed for accretion. These two examples suggest that
jets significantly contribute to the removal of angular momentum from 
accretion disks at the launching radius, both in the Class I and II phases.

Altough the previous examples suggest that jet rotation is observationally
well established, contradictions found in other sources raise the question
whether these tiny velocity shifts really trace rotation. Using optical lines
with HST/STIS, \citet{Woitas05} reported the detection of rotation in the jet 
of RW~Aur based on such velocity shifts, but with a large dispersion 
(see their Figs.~4 and 5). However,  the disk of RW~Aur was found to rotate
in the \textit{opposite} direction, based on CO 2--1 observations with the 
PdBI \citep[][]{Cabrit06}, which casts serious doubts about the interpretation
of the velocity shifts in the jet as tracing rotation. Further observations
of the jet were performed in the near-UV \citep[][]{Coffey12}. Velocity shifts
in the approaching jet lobe were found to be consistent with the disk rotation,
thus in contradiction with the velocity shifts measured in the optical.
But no velocity shift was apparent in the receding jet, and no velocity shift
was found again in the approaching jet six months later. The detection of jet 
rotation in RW~Aur has therefore not been secured yet.

\subsection{Jet rotation in the Class 0 phase}
\label{ss:jets_class0}

Evidence for jet rotation has been claimed for several Class~0 protostars. 
For instance, velocity gradients across the jet of NGC~1333~IRAS~4A2 were 
reported in SiO~1--0 with the VLA 
\citep[$HPBW \sim 1.5$--2$''$, i.e. 350--470~AU,][]{Choi11}. The 
velocity gradient has the same direction for both lobes and is seen along 8 
cuts perpendicular to the jet. In addition, the direction of rotation is the
same as for the ``disk'' \citep[][, see Sect.~\ref{ss:disks_class0}]{Choi10}.
One caveat concerning the analysis is that the velocity gradient is inferred
from simply connecting ``blobs'' in the P-V diagrams and is therefore 
somewhat subjective \citep[see Fig.~2 of][]{Choi11}. The proper motion of the
$H_2$ outflow was measured by \citet{Choi06} who derived a poloidal velocity
$v_{\rm p} \sim 70$~km~s$^{-1}$. It was assumed to be the same for the SiO 
component. The radius of the SiO jet is found to increase linearly and the 
angular speed of the outer layer to decrease with distance from the protostar
\citep[see Fig.~3 of][]{Choi11}. The specific angular momentum slightly 
increases with distance, which led the authors to suggest that the angular
momentum injection at the basis of the jet decreases with time. At least,
there seems to be no significant loss through interaction with the ambient
medium. In the framework of the magnetocentrifugal launching paradigm, the
derived poloidal and azimuthal velocities imply a launching radius 
$R_0 \sim 2$~AU (see their Fig.~5), which favors the extended disk-wind model 
over the X-wind and stellar-wind models, and a mass ejection efficiency as
traced by the SiO component 
$f_{\rm m} = \frac{\dot{M}_{\rm ej}}{\dot{M}_{\rm acc}} \sim 1\%$. The latter is
only a crude estimate and needs to be verified with a measurement of the mass
ejection rate.

\subsection{Outflow rotation}
\label{ss:outflow_rotation}

So far, we discussed claims for rotation in jets. Rotation has also been 
searched for in molecular outflows. Molecular outflows are less collimated 
than jets which typically have a full opening angle $\lesssim 10^\circ$. 
The former may consist of cloud material entrained by the latter, or material 
launched at larger radii than jets 
and maybe appearing at the stage of the first hydrostatic core already
\citep[see, e.g.][]{Machida08,Hennebelle08}. The molecular outflow of CB26, a
Class~I object in Taurus-Auriga, was mapped with the PdBI in CO 2--1 with
an angular resolution of $\sim 1.5''$ \citep[about 210~AU,][]{Launhardt09}.
It is small and well collimated, and it has the same direction as the HH objects
associated with CB26. A clear velocity gradient is detected orthogonal
to the outflow axis (see their Fig.~4). The gradient has the same orientation
as the one measured in the disk. \citet{Launhardt09} perform an empirical,
parametric modeling of the disk and outflow interferometric data, also using
constraints from the spectral energy distribution and near-infrared maps.
They find that a rotating outflow is consistent with the observations (see 
their Figs.~5 and 6), even if the exact shape of the rotation velocity profile 
across the outflow
is not well constrained. They derive a mass and total angular momentum of the 
outflow two orders of magnitude smaller than the mass and angular momentum of
the disk, and a specific angular momentum in the outflow similar to the one in 
the disk. They conclude that it will take about 1~Myr for the outflow to 
dissipate the mass and angular momentum of the disk, which is comparable to 
(although somewhat shorter than) the statistical dispersion timescale of disks 
in the Class~II phase. In the framework of the magnetocentrifugal paradigm, the 
measurements imply a launching radius $R_0 \sim 5 \pm 4$~AU, the uncertainty 
being related to the inclination \citep[][]{Cabrit09}.
Although the interpretation in terms of rotating outflow seems plausible, 
\citet{Launhardt09} do not exclude alternative interpretations, such as the
presence of two unresolved outflows or jet precession.

\subsection{Statistics}
\label{ss:jets_statistics}

If the interpretation of the velocity gradients across the jets in terms of
rotation is correct and if the magnetocentrifugal paradigm is the right one, 
then the current measurements are all consistent with the extended disk-wind
model \citep[see Fig.~3 of][ and Fig.~5 of \citeauthor{Cabrit09} 
\citeyear{Cabrit09}, and the discussions therein]{Ferreira06}. The ratios of 
mass ejection to mass accretion rates inferred from the analysis are 
$\frac{\dot{M}_{\rm ej}}{\dot{M}_{\rm acc}} < 0.1$--0.3 \citep[][]{Ferreira06}.
They are consistent with the ratios directly measured for Class~II sources
\citep[$\sim 0.1$--0.2,][]{Cabrit07}.

\begin{table}[t]
\begin{center}
\caption{Measurements of rotation in jets/outflows.}
\label{t:jets_statistics}
\vspace*{-1ex}
\begin{tabular}{lcllclll}
 \hline
 \hline
 \noalign{\smallskip}
 Jet/    & YSO & $\lambda$ & Obs. & $\Delta V$$^b$ & Bipolar & Jet/ & Ref.$^e$\\
 outflow & class & range & mode$^a$ &            & lobes$^c$ & disk$^d$ & \\
 \hline
 \noalign{\smallskip}
DG Tau   & II & VIS     & HST $\parallel$   & Y  & --      & agreem.  & 1, 2 \\
         &    & VIS     & HST $\perp$       & Y  & --      & agreem.  & 3 \\
         &    & NUV     & HST $\perp$       & Y  & --      & agreem.  & 3 \\
RW Aur   & II & VIS     & HST $\parallel$   & Y  & agreem. & disagr.  & 4, 5 \\
         &    & VIS     & HST $\perp$       & Y  & agreem. & disagr.  & 6 \\
         &    & NUV     & HST $\perp$      & Y/N & ?       & agreem.  & 7 \\
TH 28    & II & VIS     & HST $\perp$       & Y  & agreem. & --       & 6 \\
         &    & NUV     & HST $\perp$       & Y  & agreem. & --       & 3 \\
CW Tau   & II & VIS     & HST $\perp$       & Y  & --      & agreem.  & 3 \\
HH 30    & II & VIS     & HST $\perp$       & ?  & --      & inconcl. & 6, 8 \\
         &    & mm      & PdBI              & N  & --      & inconcl. & 8 \\
CB26     & I  & mm(CO)  & PdBI              & Y  & agreem. & agreem.  & 9 \\
HH 26    & I  & NIR     & VLT $\perp$       & Y  & --      & --       & 10 \\
HH 72    & I  & NIR     & VLT $\perp$       & Y  & --      & --       & 10 \\
HH 111   & I  & NIR     & GEMINI $\perp$    & N  & --      & disagr.  & 11 \\
HH 34    & I  & NIR     & GEMINI $\perp$    & U & --      & --       & 11 \\
HH 212   & 0  & NIR     & UKIRT $\parallel$ & Y  & disagr. & agreem.  & 12, 13\\
         &    & NIR     & GEMINI $\perp$    & Y  & compat. & agreem.  & 11 \\
         &    & mm(SiO) & PdBI              & Y  & disagr. & disagr.  & 14 \\
         &    & mm(SiO) & SMA               & Y  & agreem. & agreem.  & 15 \\
HH 211   & 0  & mm(SiO) & SMA               & Y  & agreem. & agreem.  & 16 \\
IRAS~4A2 & 0  & mm(SiO) & VLA               & Y  & agreem. & agreem.  & 17 \\
\hline
\noalign{\smallskip}
\end{tabular}
\end{center}
\vspace*{-1.5ex}
Notes: adapted from Table 1 of \citet{Bacciotti09}. 
$^a$ Observing mode. ``$\parallel$'' and ``$\perp$'' mean 
observations with a slit parallel or perpendicular to the jet axis.
$^b$ ``Y''/''N'' mean that a velocity gradient is/is not detected. ``U'' means
that the emission was unresolved.
$^c$ ``agreem.''/''disagr.'' means that a velocity gradient is found in both 
lobes with the same/opposite orientation.
A dash means that only one lobe was observed. 
$^d$ ``agreem.''/''disagr.'' means that the jet/outflow and the disk are 
found to rotate in the same/opposite direction. A dash means that the
direction of the disk rotation is not known. 
$^e$ References: 1: \citet{Bacciotti02}, 2: \citet{Testi02}, 
3: \citet{Coffey07}, 4: \citet{Woitas05}, 5: \citet{Cabrit06},
6: \citet{Coffey04}, 7: \citet{Coffey12}, 8: \citet{Pety06},
9: \citet{Launhardt09}, 10: \citet{Chrysostomou08}, 11: \citet{Coffey11}, 
12: \citet{Davis00}, 13: \citet{Wiseman01}, 14: \citet{Codella07},
15: \citet{Lee08}, 16: \citet{Lee07}, 17: \citet{Choi11}.
\end{table}

A summary of most measurements of rotation in jets or outflows is presented
in Table~\ref{t:jets_statistics} \citep[adapted from Table 1 of][]{Bacciotti09}.
Out of 13 Class 0, I, and II sources which were observed, 10 show a
velocity gradient perpendicular to the jet/outflow axis. For 5 of them, the 
orientation of the velocity gradient is consistent with the disk rotation,
which makes the interpretation in terms of jet rotation very plausible. 
However, in 4 cases, there is a disagreement either between the jet/outflow 
and disk directions of rotation (RW~Aur, HH~212), or between the two 
jet/outflow lobes (HH~212), or no velocity gradient is found while the high 
inclination of the system to the line of sight ($i \sim 90^\circ$) should have 
been favorable for a detection (HH~30, HH~111). These cases cast some doubts 
about the rotation interpretation. The detection of jet/outflow rotation is 
therefore not yet fully established. 

\subsection{Alternative interpretations}
\label{ss:jets_alternative}

Several alternative interpretations instead of rotation have been proposed to 
explain the velocity gradients across jets/outflows.
Beyond the Alfv\'en surface, a few AU above the disk surface, a jet may undergo
various kind of instabilities which can produce differential velocities on the
order of the sound speed ($\sim 10$~km~s$^{-1}$), similar to the velocity shifts
measured across the jets and interpreted as rotation \citep[][]{Bacciotti09}.
It is however likely that such instabilities would produce random or periodic
velocity gradients as a function of distance from the center, and not a regular
pattern as is observed in a few sources, with the same orientation in both 
lobes (e.g. TH~28, CB26, NGC1333~IRAS~4A2).

The effect of jet precession was investigated numerically by 
\citet{Cerqueira06}. They performed 3D simulations of a non-magnetic jet 
with variabilities in ejection direction (precession) and jet velocity 
(intermittence), and with or without rotation. An atomic and ionic network was 
included in order to compute line profiles. They found that only precession
(with intermittence) or 
only rotation produce very similar velocity shifts, implying that such velocity
shifts do not necessarily trace rotation.

\citet{Soker05} proposed a phenomenological model of jet interaction with a
warped disk to explain the velocity shifts seen across jets. In this model,
the jet interacts with the ambient gas above the disk surface and is more 
slowed down on the warped side of the disk than on the opposite side. In this
way, one side of the jet is slower than the other, which produces a 
velocity shift between both sides, not due to rotation. In this model, the
velocity shifts would not be seen for inclinations close to 90$^\circ$, i.e. 
a jet propagation in the plane of the sky. This would explain why no
velocity shift is seen in the jets of HH~30 or HH~111.

Finally, if the ambient medium is not uniformly distributed, its interaction
with the flow could create asymmetries in the propagation of the jet, which
would in turn produce velocity shifts between both sides of the jet 
\citep[][]{Bacciotti09}. However, persistent velocity patterns with increasing
distance from the star could \textit{a priori} not be produced in this way.

\subsection{Summary}
\label{ss:jets_summary}

The following requirements are needed to convincingly detect rotation in
jets or outflows. First of all, the kinematic signature, a velocity gradient 
perpendicular to the jet axis, has to be consistent all along the jet
length. Second, the signature should be consistent between both lobes and
with the disk rotation. Finally, the velocity profile should be spectrally
resolved. In light of this, the detection of jet rotation is statistically not 
yet secure. Doubts remain because some cases present inconsistencies between 
the putative directions of rotation of the disk and the jet. In addition, the 
absence of velocity shifts in nearly edge-on systems such as HH~30 or HH~111 
is puzzling since they should be the most favorable systems for the detection 
of jet rotation. Alternative interpretations -- jet instabilities, precession 
coupled with modulation of jet ejection, interaction of the jet with a warped 
disk, asymmetric shocking -- let open the question whether the measured
velocity shifts really probe jet rotation or not. 

Nevertheless, if the interpretation in terms of rotation is valid, then the
current constraints on the poloidal and azimuthal velocities are consistent 
with extended disk-wind models with an external launching radius of a few AU. 
The Class~II jets would seem to account for a large fraction (60--70\%) of the 
loss rate of angular momentum in disk that is required at the launching 
radius to enable accretion at that radius.

\section{Conclusions}
\label{s:conclusions}

\subsection{Reliability of rotation signatures}
\label{ss:reliability}

The previous sections introduced the methods used to probe rotation in the 
different phases of star formation. On large scales ($\sim 1$~pc), the velocity
gradients measured in molecular clouds imply a specific angular momentum
scaling as $R^{\sim 1.7}$ if they are interpreted in terms of rotation,
possibly meaning that the contraction of molecular clouds occurs with a loss
of angular momentum. However, this relation may simply be related to the 
scaling properties of turbulence, suggesting that the velocity gradients 
trace turbulence vorticity rather than pure, well-ordered rotation.
On the smaller scales of dense cores ($\sim 0.1$~pc), the correlation is 
similar, but solid-body rotation seems to be rare, and the velocity structure
of dense cores in protoclusters is relatively complex.  At a later stage, the
velocity gradients found in protostellar envelopes are often not orthogonal
to the outflow axis and the rotation signature may be ambiguous if the envelopes
are prolate rather than oblate or spherical. However, there are a few
``well-behaved'' cases for which the signature of rotation is relatively robust.
On small scales ($\lesssim 100$~AU), the existence of rotationally-supported
disks in the Class~0 phase is observationally not established yet. In the 
Class~I phase, rotationally-supported disks have been claimed to be detected
but there is a size issue compared to the Class~II disks and the former could
be ``pseudo-disks''. The most robust detections of rotation have been
made for the Class~II phase, where the kinematic structure of the detected
compact structures leaves little doubt that they are rotationally-supported
disks, i.e. structures in Keplerian rotation. The ubiquitous jets found
at all stages of star formation from the Class 0 to the Class~II phase are
expected to rotate if the current paradigm of magnetocentrifugal launching
holds. Several detections of jet rotation have been claimed, but they are
statistically not secure yet. If the interpretation of the measured velocity
gradients in terms of rotation is valid, then these detections imply a high
efficiency of angular momentum removal from the disk by the jet at a distance 
of a few AU from the protostar.

\subsection{Evolution of angular momentum}
\label{ss:j_evolution}

\begin{figure}[t]
%\centerline{\resizebox{1.0\hsize}{!}{\includegraphics[angle=0]{Figs/joverm_all_ees1209_book.eps}}}
\centerline{\resizebox{1.0\hsize}{!}{\includegraphics[angle=0]{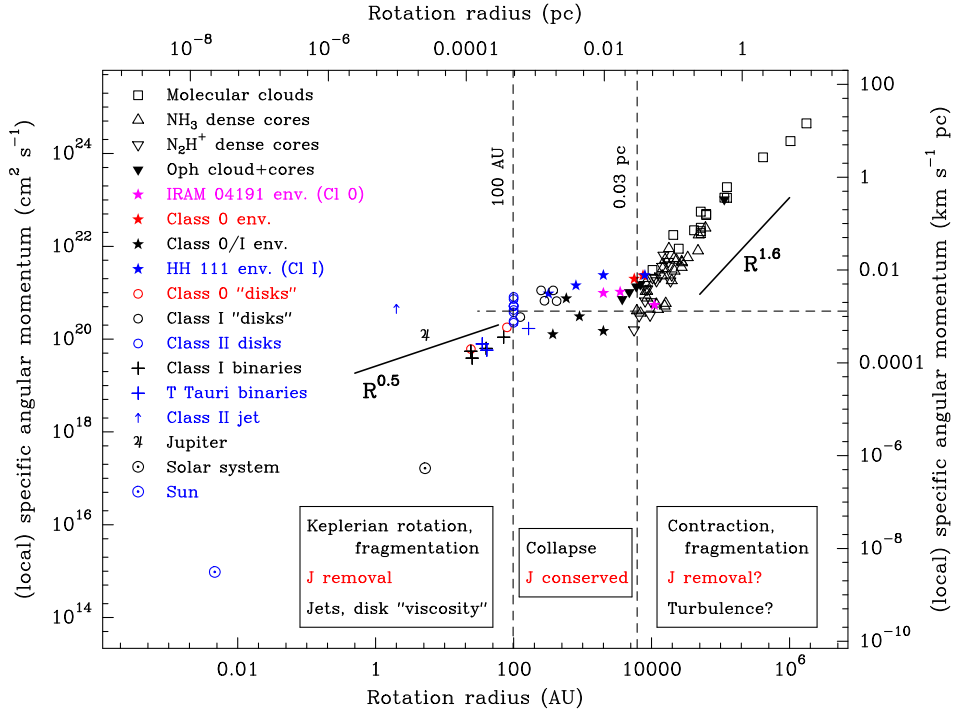}}}
\caption{Specific angular momentum as a function of radius for different 
evolutionary stages or different scales of a star forming region. The specific 
angular momentum is the local value (i.e. the product of the rotation velocity 
and the radius) for all categories except for the binaries, the solar system, 
and the Sun for which it truly corresponds to the mean angular momentum per 
unit of mass. For most categories, the radius is the radius at which the 
rotation velocity was measured or fitted and does not necessarily correspond 
to the outer radius of the object. The two solid lines are intended to guide 
the eye and are not least-square fits to the data. The vertical dashed lines 
mark the approximate position of breaks in the distribution of angular 
momentum and the horizontal one indicates the typical specific angular 
momentum during the protostellar collapse phase. References are listed in 
Table~\ref{t:jvsr}.}
\label{f:jvsr}
\end{figure}

\begin{table}[t]
\begin{center}
\caption{References for angular momenta shown in Fig.~\ref{f:jvsr}.}
\label{t:jvsr}
\vspace*{0.5ex}
\begin{tabular}{lllll}
 \hline
 \hline
 \noalign{\smallskip}
 \multicolumn{1}{c}{Category} & \multicolumn{1}{c}{Symbol} & \multicolumn{1}{c}{$\frac{J}{M}$$^{a}$} & \multicolumn{1}{c}{Corr.$^{b}$} & \multicolumn{1}{c}{Ref.$^{c}$} \\
 \hline
\noalign{\smallskip}
\multicolumn{5}{l}{\textbf{Molecular clouds and dense cores}}\\
\noalign{\smallskip}
Molecular clouds       & open square & local & no & 1 \\
NH$_3$ dense cores     & open triangle (up) & local & no & 2 \\
N$_2$H$^+$ dense cores & open triangle (down) & local & no & 3 \\
Ophiuchus cloud/cores & filled triangle & local & no & 4 \\
\noalign{\smallskip}
\multicolumn{5}{l}{\textbf{Protostellar envelopes}}\\
\noalign{\smallskip}
IRAM~04191 env. (Cl. 0) & purple star symbol & local & yes & 5--6 \\
Class 0 envelopes     & red star symbol & local & no & 7 \\
Class 0 and I envelopes & black star symbol & local & yes & 8 \\
HH~111 env. (Cl. I) & blue star symbol & local & yes & 9 \\
\noalign{\smallskip}
\multicolumn{5}{l}{\textbf{Circumstellar disks}}\\
\noalign{\smallskip}
Class 0 ``disks'' & red circle & local & yes & 10--11 \\
Class I ``disks'' & black circle & local & yes & 12--15 \\
Class II disks & blue circle & local & yes & 16 \\
\noalign{\smallskip}
\multicolumn{5}{l}{\textbf{Jets}}\\
\noalign{\smallskip}
Class II jet & blue arrow & local & yes & 17 \\
\noalign{\smallskip}
\multicolumn{5}{l}{\textbf{Young binaries}}\\
\noalign{\smallskip}
Class I binaries & black plus symbol & mean & no & 7, 18 \\
T Tauri binaries & blue plus symbol & mean & no & 7, 18 \\
\noalign{\smallskip}
\multicolumn{5}{l}{\textbf{Solar system}}\\
\noalign{\smallskip}
Jupiter & $\jupiter$ & local & yes & 19 \\
Solar system & black $\odot$ & mean & yes & 19 \\
Sun & blue $\odot$ & mean & yes & 20 \\
\hline
\noalign{\smallskip}
\end{tabular}
\end{center}
\vspace*{-1.5ex}
Notes: $^a$ Local ($v_{\rm rot} \times R_{\rm rot}$) or mean ($\frac{J}{M}$)
specific angular momentum.
$^b$ Flag indicating if the correction for inclination was applied.
$^c$ References: 1: \citet{Goldsmith85}, 2: \citet{Goodman93}, 
3: \citet{Caselli02}, 4: \citet{Belloche02b}, 5: \citet{Belloche02a}, 
6: \citet{Belloche04a}, 7: \citet{Chen07}, 8: \citet{Ohashi97}, 
9: \citet{Lee10}, 10: \citet{Choi10}, 11: \citet{Tobin12b}, 
12: \citet{Brinch07}, 
13: \citet{Lommen08}, 14: \citet{Jorgensen09}, 15: \citet{Takakuwa12}, 
16: \citet{Simon00}, 17: \citet{Bacciotti02}, 
18: Chen (2013, \textit{priv. comm.}), 19: \citet{Allen73}, 20: \citet{Pinto11}.
\end{table}

As a summary of this review, Fig.~\ref{f:jvsr} shows the distribution of 
specific angular momentum as a function of radius for different evolutionary
stages -- from molecular clouds to the Sun -- or different scales of a star 
forming region -- e.g., protostellar envelope, circumstellar disk, jet. 
References are listed in Table~\ref{t:jvsr}. This figure is partly based 
on diagrams published by \citet{Ohashi97}, \citet{Belloche02a}, and 
\citet{Chen07}. There are several caveats to this plot. First of all, for most
categories of objects, the angular momentum was derived assuming that velocity 
gradients do trace rotation, which, as was seen in the previous sections, may
not always be true. Second, a correction for inclination was applied 
for most categories but not all (see Table~\ref{t:jvsr}). Third, the plot is a 
mixture of an evolutionary diagram -- where the radius can be viewed as 
probing the contraction state of an object -- and a snapshot of the structure 
of several objects -- for which the diagram displays the angular momentum 
profile at a given time. Finally, the sample is certainly not complete and, in 
particular, upper limits for objects where no rotation was detected -- in some 
cases maybe because of an unfavorable inclination, though -- are not shown. 

Despite these caveats, the overall shape of the distribution of specific 
angular momentum as a function of radius is likely real. Three regimes can be 
distinguished. On large scales, the angular 
momentum of molecular clouds and dense cores appears to correlate well with
the radius. The specific angular momentum is reduced by nearly two orders
of magnitude over two orders of magnitude in radius. On the one hand, as 
summarized in Sect.~\ref{ss:cores_summary}, this loss of angular momentum 
could be due to magnetic braking or gravitational torques, both transfering 
angular momentum to the ambient medium, or fragmentation -- part of the 
angular momentum of the fragmenting structure being transfered into the 
orbital motion of its fragments. On the other hand, the correlation between 
``angular momentum'' (derived from velocity gradients) and radius may not be 
directly related to rotation but rather to the properties of interstellar 
turbulence and the velocity gradients may trace the behaviour of turbulence 
vorticity. More details about the physical processes at work during this phase 
can be found in the introductory chapter of Hennebelle et al. in this volume.

The second regime concerns the scales between $\sim$~100~AU and 0.03~pc (1.5 
orders of magnitude), which correspond to the scales over which a protostellar 
envelope is dynamically collapsing. In this regime, as already summarized in 
Sect.~\ref{ss:envelopes_summary}, the specific angular momentum seems to be 
relatively constant, as was first noticed by \citet{Ohashi97}. This suggests 
that processes such as magnetic braking or fragmentation are relatively 
inefficient in reducing the specific angular momentum over this range of 
scales during the protostellar collapse. Note that \citet{Yen11} discuss a
possible evolution in time of the angular momentum plateau in this regime.

Third, the few datapoints shown below $\sim$~100~AU in Fig.~\ref{f:jvsr} as 
well as the resolved structure of Class II disks (see 
Sect.~\ref{ss:disks_class2}) suggest that Keplerian rotation dominates the 
kinematics on these small scales, which implies that efficient mechanisms 
removing a large fraction of the angular momentum from the inner parts are at 
work to allow accretion to proceed. Processes contributing to the outward 
transfer of angular momentum within circumstellar disks are discussed by 
S. Fromang in this volume. In addition, the tentative detection of rotation in 
jets, if confirmed, indicates that they do extract from circumstellar disks a 
significant fraction of the angular momentum in the (small) launching region 
(see also the review of J. Ferreira 
in this volume). Fragmentation and the formation of binaries on these small 
scales can also significantly help solving the angular momentum problem by 
storing a large fraction of the angular momentum into the orbital motions of 
the stars. Systems forming single stars, like our solar system, do however not 
benefit from this mechanism.

Finally, if we extrapolate the average level of specific angular momentum at 
100~AU in circumstellar disks down to the radius of the Sun with a power-law 
consistent with Keplerian rotation ($v_{\rm rot}(R) \times R \propto R^{0.5}$), 
then we obtain a value still three orders of magnitude higher than the 
specific angular momentum of the Sun. Processes that can reduce the specific 
angular momentum of a young star during its evolution (star-disk interactions,
stellar winds) are addressed by J. Bouvier and J. Ferreira in this volume.

\subsection{Will ALMA help us?}
\label{ss:alma}

ALMA is approaching its 
completion. The early results obtained in Cycle 0 with about one third of the 
final number of antennae are already impressive and very promising for the 
future operation with the full array. With its high angular and spectral 
resolutions, ALMA will be a prime instrument to study the kinematics of star 
forming regions at very small scales with molecular line observations.

According to \citet{Schieven12},
the finest angular resolution obtained at the highest frequency ($\sim 900$~GHz)
will be about 0.005$''$ with the full array in its most extended 
configuration, i.e. 0.7~AU at 140~pc. The spectral line sensitivity 
for these extreme conditions -- $\sigma > 9000$~K in 1 min at a spectral
resolution of 0.3 km/s, or $> 400$~K in 10~h -- will not be sufficient to 
detect thermalized (non-masing) transitions, but lower-frequency 
observations are still promising: for instance, a sensitivity of 330~K 
for a spectral resolution of 2 km~s$^{-1}$ and an angular resolution of 
0.03$''$ -- 4~AU at 140~pc -- should be achievable at 150~GHz in 1 min, which 
translates into 13~K in 10~h. The capabilities of ALMA should thus allow to 
verify and firmly establish the presence of rotation in jets as well as
fully resolve rotation profiles of disks around Class 0 and I sources
and thereby significantly contribute to our understanding of the formation and 
evolution of circumstellar disks.

\vspace*{3ex}
\noindent\textit{Acknowledgments:} I am grateful to Sylvie Cabrit for 
enlightening discussions about rotation in jets. I thank Sean Andrews, 
Xuepeng Chen, Dan Clemens, Sami Dib, Stephane Guilloteau, Ralf Launhardt, 
Christopher McKee, and Markus Nielbock for discussions and/or for providing 
additional information concerning their published work. I also thank
Sylvie Cabrit and Patrick Hennebelle for their careful reading of the
manuscript. Finally, I warmly thank 
Corinne Charbonnel and Patrick Hennebelle for inviting me to the lively and 
very instructive Evry Schatzman summer school in the French alps where I
gave this lecture. 

%\begin{figure}[h]
%\centerline{\resizebox{0.30\hsize}{!}{\includegraphics[angle=270]{}}}
%\caption{}
%\label{f:}
%\end{figure}

%%-----------------------------
%%      your bibliography
%%-----------------------------

\end{document}